\documentclass[letterpaper]{emulateapj}

\usepackage{subfigure}
\usepackage{lscape}
\newcommand{\TN}{\ensuremath{T_\mathrm{e}}([\ion{N}{2}])}		%T(N II)
\newcommand{\TO}{\ensuremath{T_\mathrm{e}}([\ion{O}{3}])}		%T(O III)
\newcommand{\dens}{\ensuremath{n_\mathrm{e}}([\ion{S}{2}])}		%n(S II)
\newcommand{\dencl}{\ensuremath{n_\mathrm{e}}([\ion{Cl}{3}])}		%n(Cl III)
\newcommand{\denar}{\ensuremath{n_\mathrm{e}}([\ion{Ar}{4}])}		%n(Ar IV)
\newcommand{\n}{\ensuremath{n_\mathrm{e}}}				%ne
\newcommand{\T}{\ensuremath{T_\mathrm{e}}}				%Te
\newcommand{\cm}{\ensuremath{\mathrm{cm^{-3}}}}				%cm
\shorttitle{The iron abundance in Galactic PNe}
\shortauthors{Delgado-Inglada et al.}

\begin{document}

\title{The Iron abundance in Galactic Planetary Nebulae
\footnote{Partly based on observations made with the 2.1-m telescope at 
Observatorio Astron\'omico Nacional, San Pedro M\'artir, Mexico.}}

\author{G. Delgado-Inglada and M. Rodr\'\i guez}
\affil{Instituto Nacional de Astrof\'isica, \'Optica y Electr\'onica (INAOE), 
Apdo Postal 51 y 216, 72000 Puebla, Mexico.}

\and

\author{A. Mampaso and K. Viironen} 
\affil{Instituto de Astrof\'isica de Canarias (IAC), C/V\'ia L\'actea s/n, E38205 La Laguna, Tenerife, Spain}

\email{gloria@inaoep.mx, mrodri@inaoep.mx, amr@iac.es, kerttu@iac.es}

\begin{abstract}
We constrain the iron abundance in a sample of 33 low-ionization Galactic
planetary nebulae (PNe) using [\ion{Fe}{3}] lines and correcting for the
contribution of higher ionization states with ionization correction factors 
that take into account uncertainties in the atomic data. 
We find very low iron abundances in all the objects, suggesting that 
more than 90\% of their iron atoms are condensed onto dust grains.
This number is based on the solar iron abundance and implies a lower limit
on the dust-to-gas mass ratio, solely due to iron, of 
$M_{\rm dust}/M_{\rm gas}\geq1.3\times10^{-3}$ for our sample.
The depletion factors of different PNe cover about two orders of magnitude,
probably reflecting differences in the formation, growth, or destruction of
their dust grains. 
However, we do not find any systematic difference between the gaseous iron
abundances calculated for C-rich and O-rich PNe, suggesting similar iron
depletion efficiencies in both environments.
The iron abundances of our sample PNe are similar to those derived 
following the same procedure for a group of 10 Galactic \ion{H}{2} regions.
These high depletion factors argue for high depletion efficiencies of
refractory elements onto dust grains both in molecular clouds and 
asymptotic giant brach stars,
and low dust destruction efficiencies both in interstellar and
circumstellar ionized gas.
\end{abstract}

\keywords{planetary nebulae: general -- ISM: abundances --  dust, extinction}

\section{Introduction}
\label{intro}

The progenitors of planetary nebulae (PNe), asymptotic giant branch (AGB)
stars, have atmospheres particularly favorable for grain 
formation, and are considered the most efficient 
source of circumstellar dust \citep[and references therein]{Whittet_03,Ferrarotti_06}.
However, it is not clear yet how much dust do PNe have and whether this dust
is destroyed or modified during their lifetime.
\citet{Pottasch_84} and \citet{Lenzuni_89} studied PNe with {\it IRAS} data and found
that their derived dust-to-gas mass ratios decreased with nebular radius
(which they used as a proxy for nebular age). 
However, these results strongly depend on the poorly known distances to the
studied PNe, and were called into question by \citet{Stasinska_99}, who
derived dust-to-gas ratios using distance-independent quantities and found no
correlation with the surface brightness in H$\beta$, their proxy for
nebular age. \citeauthor{Stasinska_99} concluded that there is no evidence for
a decrease in the dust-to-gas mass ratio as PNe evolve, but
since there are many uncertainties involved, the issue is far from being
settled.

As an alternative to dust-to-gas mass ratios derived from infrared emission,
one might consider studying dust through element depletions.
Elements such as Al, Ca, Si, Ni and Fe have abundances in the 
interstellar medium (ISM) much lower than solar \citep{Morton_73, 
Morton_74}, and this is generally interpreted as due to their depletion in dust grains.
The differences in depletion factors found in different environments give
important clues on the nature of the formation and destruction mechanisms for 
dust grains  in the ISM \citep[see, e.g.,][]{Whittet_03}, and the depletion
factors in PNe can provide clues on the mechanisms that operate in ionized gas.
To study the sensitivity of depletions to environment one must choose an element
that is mostly condensed into dust grains, like those mentioned above,
since in that case the destruction of a small quantity of dust will translate
into a measurable increase of the element abundance in the gas.
However, the abundances of these elements are usually difficult to measure in
ionized gas due to the lack of suitable emission lines or atomic data, and due to
the highly uncertain corrections for unobserved ions.
These problems, combined with the wide spread in the degrees of ionization, and
hence of ionization states, found in PNe, imply that the depletion factors
derived so far for PNe use different ions and ionization correction factors
(ICFs) and are not only uncertain, but also difficult to compare between them.
The published values for the abundances of refractory elements in PNe cover the
ranges: 1/6--1/300 the solar abundance for Ca, 
1/2--1/350 for Al, 
1/3--1/300 for Fe, 
near solar to 1/10 solar for Mg, 
and near solar to 1/20 solar for Si
\citep{Shields_75, Garstang_78, Shields_78,Pequignot_80,Aller_81,
Shields_81,Aller_83,Beckwith_84,Pwa_86,Clegg_87a, Clegg_87b, Middlemass_90,
Keyes_90, Kingdon_95,Pottasch_99,Perinotto_99,Casassus_00,Pottasch_01,
Pottasch_02,Pottasch_03,Liu_04a,Pottasch_05,Sterling_05,Georgiev_06,
Likkel_06, Pottasch_07a,Pottasch_07b,Pottasch_08}.

The problem with the determination of depletion factors in ionized gas is
somewhat alleviated in the case of \ion{H}{2} regions, where the range of
degrees of ionization is much smaller. However, dust grains in PNe
and \ion{H}{2} regions are likely to have very different characteristics.
The dust grains present in PNe formed in the cool atmospheres
of the progenitor stars, whereas those grains now present in \ion{H}{2} regions
were located before in the associated molecular clouds and can be considered
processed interstellar dust grains.
Therefore, it is important to perform a homogeneous study of depletion 
factors in a sample of PNe, and especially so if the results can also be compared 
with those found in \ion{H}{2} regions.
Differences in depletion factors can provide much information on the 
efficiency of dust formation and destruction processes.

Of all the refractory elements we mentioned above, Fe has the strongest lines 
in the visible range of the spectrum.
Furthermore, since most of the Fe atoms are condensed into dust grains and
since the cosmic abundance of Fe is relatively high, this element is an
important contributor to the mass of refractory dust grains \citep{Sofia_94}, 
and the Fe gaseous abundance will probably reflect the abundance of refractory
elements in dust. These reasons make Fe a good choice to study depletion
factors in ionized gas. 

The comparison between the results derived in PNe and \ion{H}{2} regions will
be more meaningful if the sample of PNe is restricted to objects with a low
degree of ionization, because in that case the same ions need to be considered
in the abundance determination for both types of objects. In \ion{H}{2}
regions, Fe will be mostly found in three ionization states: Fe$^{+}$, Fe$^{++}$, and
Fe$^{+3}$. Fe$^{+}$ has a low ionization potential and its contribution to the 
total abundance is often negligible \citep{Rodriguez_02}.
On the other hand, [\ion{Fe}{4}] lines are weak and more difficult to measure 
than [\ion{Fe}{3}] lines. Hence, Fe abundances are usually calculated 
from Fe$^{++}$ abundances and an ICF derived from photoionization 
models.
However, for the handful of objects in which [\ion{Fe}{4}] lines have been
measured, this ICF can be compared with that implied by the derived
Fe$^{+3}$ abundances, and a discrepancy has been found between them \citep[and
references therein]{Rodriguez_03}.
\citet{Rodriguez_05} determined what changes in all the atomic data involved in
the calculations would explain this discrepancy: (1) a decrease in the
collision strengths for Fe$^{+3}$ by factors of 2--3, (2) an increase in the
collision strengths for Fe$^{++}$ by factors of 2--3, or (3) an increase in the
total recombination  coefficient or the rate of the charge-exchange reaction
with  H$^0$ for Fe$^{+3}$ by a factor of $\sim10$.
\citet{Rodriguez_05} argued that the three explanations are equally plausible, 
and derived two different ICFs: 
\begin{equation}\label{eq1}
\frac{\mbox{Fe}}{\mbox{O}}=0.9\left(\frac{\mbox{O}^{+}}{\mbox{O}^{++}}\right)^{0.08}\frac{\mbox{Fe}^{++}}{\mbox{O}^{+}}
\end{equation}
\begin{equation}\label{eq2}
\frac{\mbox{Fe}}{\mbox{O}}=1.1\left(\frac{\mbox{O}^{+}}{\mbox{O}^{++}}\right)^{0.58}\frac{\mbox{Fe}^{++}}{\mbox{O}^{+}}
\end{equation}
Equation~(\ref{eq1}) is based on photoionization models that use the 
state-of-the-art values for the atomic data relevant to the problem, and 
Equation~(\ref{eq2}) is derived from those objects with measurements of 
[\ion{Fe}{4}] lines \citep[see][]{Rodriguez_05}.
Equation~(\ref{eq2}) should be replaced by
\begin{equation}\label{eq3}
\frac{\rm {Fe}}{\rm {O}}=\frac{\mbox{Fe}^+ + \mbox{Fe}^{++}}{\mbox{O}^+}
\end{equation}
for those objects with $\log(\mbox{O}^+/\mbox{O}^{++})\geq-0.1$, since
$\mbox{Fe}^{++}$ and $\mbox{O}^+$ will then dominate the total abundances of Fe
and O. If the discrepancy were completely due to errors in the
collision strengths  for Fe$^{+3}$, good values of the Fe abundance could be
obtained from the  ICF of Equation~(\ref{eq1}). If the collision strengths for
Fe$^{++}$ were the  ones to blame, the correct abundance would be the previous
value lowered  by $\sim 0.3$~dex. Finally, if the models predictions were
wrong  -- due to errors in the total recombination coefficient or the rate of
the charge-exchange reaction -- the ICF of equation~(\ref{eq2}) would give the
best values of the Fe abundance. The discrepancy is probably due to some
combination of the aforementioned causes, therefore the errors required in any
of the atomic data are likely to be lower than those considered above, and the
values of the Fe abundance will consequently be intermediate between the
extreme values obtained with the ICF scheme described above.

The three ionization correction schemes give values of the Fe abundance that
can be similar or differ by more than a factor of 10, but since these schemes
involve drastic changes in the atomic data involved in the abundance
calculation, the extreme values of the Fe abundance implied by them can be used
to constrain the true values of the Fe abundances in the gas.
\citet{Rodriguez_05} followed this procedure and found that the two Galactic 
\ion{H}{2} regions and the four Galactic PNe of their sample have less than 5\%
of their Fe atoms in the gas phase.
In this paper, we apply this procedure to
constrain the Fe abundances in a sample of 33 low-ionization Galactic PNe,
and compare the results  with the values obtained for 10 Galactic \ion{H}{2}
regions.

\section{The sample}
\label{sample}

In order to perform the analysis described above, we need to select
a sample of low-ionization PNe where Fe$^{+}$, Fe$^{++}$, and
Fe$^{+3}$ (the latter with ionization potential IP = 54.8 eV) are the main 
ionization states of iron and O$^{+}$, O$^{++}$ (the latter with IP = 54.9 eV)
are the main ionization states of oxygen. We considered a group of PNe 
that have determinations of the O$^{+3}$ abundance \citep{Liu_04a,Tsamis_03},
and found that those with
$I(\mbox{\ion{He}{2}~}\lambda4686)/I(\mbox{H}\beta)\lesssim0.3$
have less than 10\% of their O abundance in this ionization state.
Hence, we established this as the condition that our sample PNe
should satisfy.

The initial sample consists of 28 low-ionization PNe, 23 of them selected 
from the literature because their published spectra have all the 
lines we need to calculate physical conditions and the Fe$^{++}$ 
and O ionic abundances. Three of these PNe do not have measurements 
of [\ion{Fe}{3}] lines, but do have measurements of weak, nearby 
recombination lines which can be used to calculate upper limits to 
the Fe$^{++}$ and Fe abundances.
The other 5 PNe of the sample were observed in the 2.1-m telescope 
at Observatorio Astron\'omico Nacional (San Pedro M\'artir, Mexico).

Our initial sample does not contain PNe with high electron 
densities ($\n\gtrsim25,000~\cm$) because it is difficult to obtain 
good estimates of the physical conditions in these objects. 
However, we have included 5 additional PNe with \n\ $> 25,000 \cm$, and 
performed a special analysis, as described in Section 6.

\section{Observations and data reduction}

Long-slit spectra covering the wavelength ranges $\lambda\lambda$3600--5700
and  $\lambda\lambda$5350--7500 were obtained with the Boller \& Chivens
spectrograph and the SITe3 CCD detector in the 2.1-m telescope at Observatorio 
Astron\'omico Nacional (San Pedro M\'artir, Mexico).
The spectral ranges were covered 
with a spectral resolution of $\sim4$ \AA\ using a 600 lines $\mbox{mm}^{-1}$ 
grating at two different angles, and a slit width of 2\arcsec. 
The observed objects, the positions and position angles (P.A.) of the slit, 
and the exposure times are listed in Table~\ref{tabla_1}.
The slit was positioned at the center of the nebulae with the exception of 
JnEr~1, where the slit was placed at the NW condensation.
Bias frames, twilight and tungsten flat-field exposures, wavelength
calibrations, and exposures of standard stars were taken each night.
The angular diameters of IC~4593, NGC~2392, and NGC~6210 
are smaller than the slit length and suitable sky 
windows could be selected on either side of the nebular emission. 
For NGC~3587 and JnEr~1, sky spectra were obtained near the objects after the
nebular exposures. The spectra were reduced 
using the {\sc IRAF}\footnote{{\sc IRAF} is distributed by the National 
Optical Astronomy Observatories, which are operated by the Association 
of Universities for Research in Astronomy, Inc., under cooperative 
agreement with the National Science Foundation.} reduction package 
and following the standard procedures for long-slit reductions. After 
the bias subtraction, flat-field correction, and wavelength calibration, 
the images were flux calibrated with the standard stars Feige~34, Feige~56 
and G191B2B. We subtracted the sky (after scaling it by factors of 
0.5--1.5 to obtain the best cancelation in those cases where the sky spectra
were observed separately), and removed cosmic rays by 
the combination of different exposures.
Finally, one-dimensional spectra were extracted.

\begin{deluxetable*}{lcccll}
\tabletypesize{\small}
\tablecaption{Journal of observations\label{tabla_1}}
\tablewidth{0pt}
\tablehead{
\colhead{Object} & \colhead{$\alpha$ (2000)} & \colhead{$\delta$ (2000)} &
\colhead{P.A.} & \colhead{Date\tablenotemark{a}} &
\colhead{Exposure times\tablenotemark{a}}\\
 & \colhead{(hh mm ss )} & \colhead{(\degr\ \arcmin\ \arcsec)}  &
 \colhead{(\degr)} & \colhead{} & \colhead{(s)}
 }
\startdata
IC~4593  & 16 11 44.54 & $+$12 04 17.06 & 0  & 2006 Jan 28, 29 & 10, 3$\times$30, 60, 29$\times$120\\
         &             &                &    & 2006 Jan 27     & 30, 4$\times$40, 2$\times$100, 5$\times$300\\
JnEr~1  & 07 58 19.00 & $+$53 25 17.00 & 90 & 2006 Jan 28, 29 & 12$\times$1200\\
         &             &                &    & 2006 Jan 27     & 4$\times$1200\\
NGC~2392 & 07 29 10.77 & $+$20 54 42.49 & 65 & 2007 Jan 22, 24 & 6$\times$60, 17$\times$120\\
         &             &                &    & 2007 Jan 23     & 60, 12$\times$120\\
NGC~3587 & 11 14 47.73 & $+$55 01 08.50 & 55 & 2006 Jan 26     & 5$\times$1200\\
         &             &                &    & 2006 Jan 25     & 5$\times$1200\\
NGC~6210 & 16 44 29.49 & $+$23 47 59.68 & 90 & 2006 Jan 28, 29 & 3$\times$10, 20$\times$60\\
         &             &                &    & 2006 Jan 27     & 3, 19$\times$15
\enddata
\tablenotetext{a}{The first entry corresponds to the blue range
$\lambda\lambda$3600--5700 and the second one corresponds to the red range
$\lambda\lambda$5350--7500.}
\end{deluxetable*}

Line intensities were measured by integrating between two given 
limits above a continuum around each line estimated by eye. In 
the cases of line blending, a multiple Gaussian profile-fitting
procedure was applied to obtain the intensity of each individual line.
These measurements were made with the {\sc SPLOT} routine of the {\sc IRAF}
package. The line intensities were first normalized to the brightest 
\ion{H}{1} line appearing in the same spectral range: $\mbox{H}\beta$ 
for the blue range and $\mbox{H}\alpha$ for the red range. 
These line ratios were corrected for extinction using the extinction law of
\citet{Cardelli_89} with a total to selective extinction ratio
$R_{V}=A(V)/E(B-V)=3.1$, the mean value for the diffuse interstellar medium.
The logarithmic extinction $c(\rm{H}\beta)$ was calculated 
from the comparison between the observed and theoretical ratio 
$I(\mbox{H}\beta)/I(\mbox{H}\gamma)$ for typical physical conditions 
\citep{Storey_95}: $\T\ = 10,000$~K and \n\ = 100, 5000 
or 10,000 \cm\ depending on the PN.
Dereddened intensities  were obtained by multiplying the observed intensity
ratios by the factor $10^c(\rm{H}\beta)\,\rm{f}(\lambda)$, where $f(\lambda)$ comes
from the extinction law. 
Columns 1 and 2 in Table~\ref{tabla_2} show the laboratory and observed 
wavelengths, Column 3 shows the line identifications, and Columns 4 and 5 
contain the observed [$I_{\rm ob}(\lambda)$] and dereddened [$I(\lambda$)] 
line intensities, normalized with respect to $I(\mbox{H}\beta) = 100$.
The intensities of lines in the red range were normalized with respect to
$I(\mbox{H}\beta)$ using the theoretical value of the ratio
$I(\mbox{H}\alpha)/I(\mbox{H}\beta)$ and the value derived for c(H$\beta$). 
The logarithmic extinction $c(\rm{H}\beta)$, the observed and dereddened intensity 
of H$\beta$, and the extraction window are also given for each object in 
Table~\ref{tabla_2}.

\setcounter{table}{1}
\begin{table}
\begin{minipage}{75mm}
\centering \caption{Observed and reddening-corrected line ratios with respect to 
$I(\mbox{H}\beta) = 100$.\label{tabla_2}}
\begin{tabular}{c@{\hspace{2.8mm}}c@{\hspace{2.8mm}}c@{\hspace{1.8mm}}c@{\hspace{2.8mm}}c@{\hspace{2.8mm}}}
\noalign{\hrule} \noalign{\vskip2pt} \noalign{\hrule} 
$\lambda$(${\rm \AA}$)& $\lambda_{\rm ob}$(${\rm \AA}$) & Ion & $I_{\rm ob}(\lambda)$ & $I(\lambda)$\\
\noalign{\vskip3pt} \noalign{\hrule} \noalign{\vskip3pt}
\multicolumn{5}{c}{\normalsize{IC~4593}}\\
\noalign{\vskip3pt} \noalign{\hrule} \noalign{\vskip3pt}
3703.85 & 3705.02 & H\,16           & 1.50$\pm$0.45  & 1.79$\pm$0.58\\
3705.02 & *       & \ion{He}{1}     & *              &   *\\
3711.97 & 3712.73 & H\,15           & 1.21$\pm$0.11  & 1.45$\pm$0.22\\
3726.03 & 3727.28 & [\ion{O}{2}]    & 42.07$\pm$2.4  & 50.3$\pm$6.6 \\
3728.82 & *       & [\ion{O}{2}]    & *  & *\\
3770.63 & 3770.79 & H\,11           & 3.64$\pm$0.23 & 4.33$\pm$0.57\\
3797.90 & 3797.87 & H\,10           & 2.85$\pm$0.18 & 3.38$\pm$0.44\\
3819.61 & 3819.72 & \ion{He}{1}     & 1.01$\pm$0.10 & 1.20$\pm$0.18\\ 
3835.39 & 3835.63 & H\,9            & 4.50$\pm$0.27 & 5.31$\pm$0.66\\
3868.75 & 3869.01 & [\ion{Ne}{3}]   & 27.7$\pm$1.6  & 32.6$\pm$4.0\\
3888.65 & 3889.07 & \ion{He}{1}     & 16.90$\pm$0.96 & 19.8$\pm$2.4\\ 
3889.05 & *       & H\,8            & * & *\\
3967.46 & 3969.15 & [\ion{Ne}{3}]   & 21.8$\pm$1.2 & 25.2$\pm$2.9\\
3970.07 & *       & H\,7            & * & *\\
4009.22 & 4009.32 & \ion{He}{1}     & 0.333$\pm$0.089 & 0.38$\pm$0.11\\
4026.08 & 4026.58 & \ion{N}{2}      & 1.47$\pm$0.11 & 1.69$\pm$0.20\\
4026.21 & *       & \ion{He}{1}     & *  &  \\
4101.74 & 4101.95 & H\,6            & 22.4$\pm$1.3 & 25.4$\pm$2.6\\
4132.80 & 4133.61 & \ion{O}{2}      & 0.140$\pm$0.082 & 0.158$\pm$0.094 \\
4143.76 & 4144.24 & He I            & 0.339$\pm$0.094 & 0.38$\pm$0.11\\   
4153.30 & 4154.46 & O II            & 0.295$\pm$0.093 & 0.33$\pm$0.11\\
4156.53 & *       & O II          & * & *\\
4267.15 & 4267.65 & C II            & 0.416$\pm$0.095   &  0.46$\pm$0.11\\   
4340.47 & 4340.71 & H\,5            & 43.0$\pm$2.4      &  46.9$\pm$3.8\\
4363.21 & 4363.76 & [O III]         & 1.83$\pm$0.14     &  1.99$\pm$0.19\\   
4387.93 & 4388.48 &  He I           & 0.542$\pm$0.088   &  0.59$\pm$0.10\\   
4471.49 & 4471.69 &  He I           & 4.74$\pm$0.27     &  5.06$\pm$0.36\\   
4634.14 & 4634.90 &  N III          & 0.713$\pm$0.057   &  0.740$\pm$0.061\\   
4640.64 & 4641.40 &  O II           & 0.776$\pm$0.059   &  0.804$\pm$0.064\\
4641.81 & *       &  O II           & * & *\\
4641.84 & *       &  N III          & * & *\\
4643.31 & *       &  N II           & * & *\\
4647.42 & 4648.18 &  C III          & 0.739$\pm$0.058   &  0.764$\pm$0.062\\ 
4649.13 & *       &  O II           & * & *\\
4650.25 & 4651.26 &  C III          & 0.477$\pm$0.048   &  0.493$\pm$0.051\\   
4650.84 & *       &  O II           & * & *\\
4658.10 & 4659.02 &  [Fe III]       & 0.692$\pm$0.056   &  0.714$\pm$0.060\\ 
4661.63 & 4662.44 &  O II           & 0.054$\pm$0.039   &  0.056$\pm$0.040\\   
4685.68 & 4685.03 &  He II          & 0.418$\pm$0.052   &  0.430$\pm$0.054\\   
4701.62 & 4702.22 &  [Fe III]       & 0.124$\pm$0.042   &  0.127$\pm$0.043\\   
4711.37 & 4711.37 &  [Ar IV]        & 0.511$\pm$0.049   &  0.523$\pm$0.051\\  
4713.17 & *       &  He I           & * & *\\ 
4861.33 & 4861.27 & H\,4            & 100.0$\pm$5.7     & 100.0$\pm$5.7\\
4881.11 & 4880.92 & [Fe III]       & 0.118$\pm$0.039   & 0.118$\pm$0.039\\   
4890.86 & 4889.70 &  [Fe II]       & 0.058$\pm$0.036   & 0.057$\pm$0.036\\
        & 4891.72 & O II           & * & *\\ 
4906.83 & 4905.34 & [Fe II]        & 0.104$\pm$0.047   & 0.103$\pm$0.046\\  
 *      & 4907.03 &  O II          &  * & *\\ 
4921.93 & 4922.00 & He I           & 1.444$\pm$0.097   & 1.431$\pm$0.096\\   
4958.91 & 4958.77 & [O III]        & 197$\pm$11        & 194$\pm$11\\
5006.84 & 5006.67 & [O III]        & 555$\pm$31        & 544$\pm$32\\
5015.68 & 5015.99 & He I           & 1.043$\pm$0.066   & 1.020$\pm$0.066\\
5047.74 & 5047.65 & He I           & 0.184$\pm$0.031   & 0.179$\pm$0.031\\
5191.82 & 5191.88 & [Ar III]       & 0.086$\pm$0.032   & 0.072$\pm$0.030\\ 
5270.40 & 5268.77 & [Fe III]       & 0.338$\pm$0.047   & 0.320$\pm$0.046\\ 
5517.66 & 5516.59 & [Cl III]       & 0.416$\pm$0.049   & 0.384$\pm$0.048\\
5537.60 & 5536.91 & [Cl III]       & 0.340$\pm$0.044   & 0.313$\pm$0.044\\
5666.63 & 5666.31 &  N II          & 0.110$\pm$0.032   & 0.108$\pm$0.032\\
5754.60 & 5754.84 & [N II]         & 0.146$\pm$0.035   & 0.143$\pm$0.034\\
5875.66 & 5876.27 & He I           & 15.86$\pm$0.80    & 15.30$\pm$0.84\\
6300.34 & 6299.44 & [O I]          & 0.146$\pm$0.047   & 0.137$\pm$0.044\\
6312.10 & 6312.66 & [S III]        & 0.816$\pm$0.067   & 0.761$\pm$0.060\\
6548.10 & 6548.62 & [N II]         & 3.91$\pm$0.20     & 3.58$\pm$0.15\\
6562.77 & 6563.02 & H\,3           & 100.0$\pm$4.2     & 286$\pm$12\\
6583.50 & 6583.67 & [N II]         & 11.39$\pm$0.57    & 10.41$\pm$0.44\\
6678.16 & 6678.26 & He I           & 4.59$\pm$0.23     & 4.17$\pm$0.18\\
6716.44 & 6716.58 & [S II]         & 0.667$\pm$0.041   & 0.604$\pm$0.034\\
6730.82 & 6730.96 & [S II]         & 0.930$\pm$0.052   & 0.841$\pm$0.042\\
7065.25 & 7064.63 & He I           & 4.48$\pm$0.23     & 3.95$\pm$0.20\\
7135.80 & 7135.10 & [Ar III]       & 10.78$\pm$0.54    & 9.47$\pm$0.48\\
7280.76 & 7280.76 &  He I          & 0.848$\pm$0.049   & 0.737$\pm$0.046\\
\noalign{\vskip 3pt} \noalign{\hrule} \noalign{\vskip3pt}
\end{tabular}   
\end{minipage}  
\end{table}

\setcounter{table}{1}
\begin{table}
\begin{minipage}{75mm}
\centering \caption{{\it --continued}}
\begin{tabular}{c@{\hspace{2.8mm}}c@{\hspace{2.8mm}}c@{\hspace{1.8mm}}c@{\hspace{2.8mm}}c@{\hspace{2.8mm}}}
\noalign{\hrule} \noalign{\vskip2pt} \noalign{\hrule} 
$\lambda$(${\rm \AA}$)& $\lambda_{\rm ob}$(${\rm \AA}$) & Ion & $I_{\rm ob}(\lambda)$ & $I(\lambda)$\\
\noalign{\vskip3pt} \noalign{\hrule} \noalign{\vskip3pt}
\multicolumn{5}{c}{\normalsize{IC~4593 (cont.)}}\\
\noalign{\vskip3pt} \noalign{\hrule} \noalign{\vskip3pt}
7318.92 & 7319.09 &  [O II]        & 1.539$\pm$0.082   & 1.333$\pm$0.079\\
7329.67 & 7329.85 &  [O II]        & 1.201$\pm$0.065   & 1.040$\pm$0.063\\
\\
\multicolumn{5}{l}{\footnotesize{c(H$\beta\,)=\,0.24\pm0.16$}}\\
\multicolumn{5}{l}{\footnotesize{$I_{\rm ob}\ (\mbox{H}\beta)\,=\,4.462\times10^{-12}\,\mbox{erg}\ \mbox{cm}^{-2}\ \mbox{s}^{-1}$}}\\
\multicolumn{5}{l}{\footnotesize{$I(\mbox{H}\beta)\,=\,7.754\times10^{-12}\,\mbox{erg}\ \mbox{cm}^{-2}\ \mbox{s}^{-1}$}} \\
\multicolumn{5}{l}{\footnotesize{Extraction window = $2\arcsec\times15\arcsec$}} \\
\noalign{\vskip3pt} \noalign{\hrule} \noalign{\vskip3pt}
\multicolumn{5}{c}{\normalsize{JnEr~1}}\\
\noalign{\vskip3pt} \noalign{\hrule} \noalign{\vskip3pt}
3711.97 & 3710.54 & H 15      & 117$\pm$25 & 151$\pm$48\\
3726.03 & 3727.29 & [O II]    & 608$\pm$54 & 783$\pm$193\\
3728.82 & *       & [O II]    &    *       &    *\\ 
3734.37 & 3734.05 & H 13      & 41$\pm$13  & 53$\pm$21\\
3756.10 & 3758.73 & He I     & 28.7$\pm$7.7   & 37$\pm$13\\
3797.90 & 3798.16 & H 10      & 29.8$\pm$8.8   & 38$\pm$14\\
3868.75 & 3868.09 & [Ne III]  & 98$\pm$13  & 122$\pm$30\\
3967.46 & 3967.98 & [Ne III]  & 39$\pm$11  & 48$\pm$16\\
3970.07 & *       & H 7       &    *       &   *\\ 
4083.90 & 4083.61 & O II      & 16.6$\pm$3.8   & 20.0$\pm$5.6\\
4085.11 & *       & O II      &     *       &   *\\ 
4101.74 & 4102.56 & H\,6 & 34.0$\pm$5.4 & 40.7$\pm$9.3\\
4340.63 & 4340.60 & H\,5 & 41.4$\pm$3.4 & 46.8$\pm$6.5\\
4363.21 & 4361.65 & [O III]   &  5.1$\pm$1.3 &  5.7$\pm$1.6\\
4471.50 & 4471.77 & He I      &  9.7$\pm$1.8 & 10.6$\pm$2.2\\
4487.72 & 4487.61 & O II      &  2.1$\pm$1.2 & 2.3$\pm$1.3\\
4488.20 & *       & O II      &     *       &   *\\ 
4489.49 & *       & O II      &     *       &   *\\ 
4609.44 & 4610.27 & O II      & 2.81$\pm$0.67 & 2.98$\pm$0.73\\
4610.20 & *       & O II      &     *       &   *\\
4658.10 & 4659.44 & [Fe III]  & 2.34$\pm$0.75 & 2.45$\pm$0.79\\
4661.63 & *       & O II      &     *       &   *\\
4685.68 & 4685.80 & He II     & 20.1$\pm$1.6 & 20.9$\pm$1.8\\
4711.37 & 4713.69 & He I      &  2.23$\pm$0.84 & 2.31$\pm$0.87\\
4713.17 & *       & [Ar IV]   &     *        &    *\\  
4740.17 & 4740.05 & [Ar IV]   &  5.4$\pm$1.1 & 5.6$\pm$1.1\\ 
4861.33 & 4861.00 & H\,4      & 100.0$\pm$6.3   & 100.0$\pm$6.3\\ 
4958.91 & 4958.51 & [O III]   & 159.9$\pm$9.7   &157$\pm$10\\ 
5006.84 & 5006.36 & [O III]   & 456$\pm$27      & 443$\pm$29\\ 
5197.90 & 5198.36 & [N I]     & 11.60$\pm$0.89  & 10.9$\pm$1.0\\ 
5200.26 & *       & [N I]     &     *           &   *\\
5411.52 & 5411.90 & He II     & 2.88$\pm$0.54   & 3.22$\pm$0.72\\ 
5537.60 & 5537.27 & [Cl III]  & 0.75$\pm$0.37   & 0.66$\pm$0.33\\ 
5754.60 & 5754.17 & [N II]    & 8.98$\pm$0.88   & 9.6$\pm$1.2\\ 
5875.66 & 5875.66 & He II     & 27.0$\pm$1.7   & 28.3$\pm$2.5\\ 
6300.34 & 6302.66 &  [O I]    & 25.9$\pm$2.0   & 26.0$\pm$2.0\\ 
6312.10 & 6309.84 &  [S III]  & 5.98$\pm$1.0   & 6.0$\pm$1.0\\ 
6310.80 & *       & He II     &     *          &   *\\
6363.78 & 6366.14 &  [O II]   & 9.3$\pm$1.1    & 9.2$\pm$1.1\\ 
6548.10 & 6547.97 &  [N II]   & 222$\pm$12     & 217$\pm$9.7 \\ 
6562.77 & 6562.80 &  H\,3     & 294$\pm$16     & 286$\pm$13\\ 
6583.50 & 6583.36 & [N II]    & 682$\pm$37     & 664$\pm$29\\ 
6678.16 & 6678.16 &  He I     & 7.82$\pm$0.88  & 7.54$\pm$0.82\\
6716.44 & 6716.49 &  [S II]   & 28.5$\pm$1.8   & 27.3$\pm$1.6\\ 
6730.82 & 6731.01 &  [S II]   & 22.5$\pm$1.6   & 21.6$\pm$1.4\\ 
7065.25 & 7063.95 &  He I     & 4.4$\pm$1.1    & 4.1$\pm$1.0 \\ 
7135.64 & 7135.02 &  [Ar III] & 23$\pm$2.2     & 21.3$\pm$2.3 \\ 
7319.99 & 7318.65 &  [O III]  & 23$\pm$11      & 21.1$\pm$9.6\\ 
7330.73 & 7331.29 &  [O III]  & 12.9$\pm$8.9 & 11.6$\pm$8.0\\
\\
\multicolumn{5}{l}{\footnotesize{c(H$\beta\,)=\,0.34\pm0.31$}}\\
\multicolumn{5}{l}{\footnotesize{$I_{\rm ob}\ (\mbox{H}\beta)\,=\,3.663\times10^{-14}\,\mbox{erg}\ \mbox{cm}^{-2}\ \mbox{s}^{-1}$}}\\
\multicolumn{5}{l}{\footnotesize{$I(\mbox{H}\beta)\,=\,8.014\times10^{-14}\,\mbox{erg}\ \mbox{cm}^{-2}\ \mbox{s}^{-1}$}}\\
\multicolumn{5}{l}{\footnotesize{Extraction window = $2\arcsec\times42\arcsec$}} \\
\noalign{\vskip3pt}\noalign{\hrule} \noalign{\vskip3pt}
\multicolumn{5}{c}{\normalsize{NGC~2392}}\\
\noalign{\vskip3pt} \noalign{\hrule} \noalign{\vskip3pt}
3726.03 & 3728.98 & [O II]   & 105.8$\pm$6.1     & 135$\pm$28\\
3728.82 & *       & [O II]   & *     &           *\\
3835.39 & 3837.09 & H\,9     & 2.31$\pm$0.19     & 2.90$\pm$0.59\\
3868.75 & 3870.42 & [Ne III] & 95.3$\pm$5.4      & 119$\pm$22\\      
3888.65 & 3890.44 & He I     & 13.69$\pm$0.80    & 17.0$\pm$3.2\\
3889.05 & *       & H\,8     &          *        & * \\ 
\noalign{\vskip 3pt} \noalign{\hrule} \noalign{\vskip3pt}
\end{tabular}   
\end{minipage}  
\end{table}

\setcounter{table}{1}
\begin{table}
\begin{minipage}{75mm}
\centering \caption{{\it --continued}}
\begin{tabular}{c@{\hspace{2.8mm}}c@{\hspace{2.8mm}}c@{\hspace{1.8mm}}c@{\hspace{2.8mm}}c@{\hspace{2.8mm}}}
\noalign{\hrule} \noalign{\vskip2pt} \noalign{\hrule} 
$\lambda$(${\rm \AA}$)& $\lambda_{\rm ob}$(${\rm \AA}$) & Ion & $I_{\rm ob}(\lambda)$ & $I(\lambda)$\\
\noalign{\vskip3pt} \noalign{\hrule} \noalign{\vskip3pt}
\multicolumn{5}{c}{\normalsize{NGC~2392 (cont.)}}\\
\noalign{\vskip3pt} \noalign{\hrule} \noalign{\vskip3pt}
3967.46 & 3969.71 & [Ne III] & 34.3$\pm$2.0      & 42.06$\pm$7.4\\
3970.07 &   *     &  H\,7    & * & *\\
4068.60 & 4070.93 & [S II]   & 1.89$\pm$0.18     & 2.27$\pm$0.40\\
4101.74 & 4103.09 & H\,6     & 21.3$\pm$1.2      &  25.3$\pm$3.9\\
4120.84 & 4119.10 & He I     & 1.29$\pm$0.19     & 1.53$\pm$0.31\\
4121.46 & *       & O II     &       *           &          *\\  
4143.76 & 4145.07 & He I     & 0.72$\pm$0.17     & 0.84$\pm$0.23\\ 
4227.74 & 4229.36 & N II     & 0.38$\pm$0.14     & 0.44$\pm$0.17\\
4340.47 & 4341.71 & H\,5     & 41.6$\pm$2.4      & 46.9$\pm$5.3\\
4363.21 & 4364.54 & [O III]  & 16.74$\pm$0.95    & 18.8$\pm$2.0\\
4465.41 & 4467.08 & O II     & 0.772$\pm$0.077   & 0.84$\pm$0.10\\
4466.42 & *       & O II     &       *           &          *\\ 
4471.49 & 4472.94 &  He I    & 2.70$\pm$0.17     & 2.95$\pm$0.28\\
4518.15 & 4516.85 &  N III   & 1.51$\pm$0.11     & 1.63$\pm$0.16\\
4638.86 & 4637.25 &  N III   & 3.55$\pm$0.21     & 3.72$\pm$0.26\\
4641.81 & 4643.39 &  O II    & 2.02$\pm$0.13     & 2.12$\pm$0.16\\
4641.84 & *       &  N III   & * & *\\
4643.08 & *       &  N II    & * & *\\
4658.10 & 4659.23 &  [Fe III]& 2.33$\pm$0.15     & 2.44$\pm$0.18\\ 
4676.24 & 4678.39 &  O II    & 0.650$\pm$0.070   &  0.676$\pm$0.076\\
4685.68 & 4686.99 &  He II   & 30.8$\pm$1.8      &  32.0$\pm$2.1\\
4701.62 & 4703.67 &  [Fe III]& 0.515$\pm$0.069   &  0.533$\pm$0.073\\
4711.37 & 4713.07 &  [Ar IV] & 1.96$\pm$0.13     &  2.03$\pm$0.15\\  
4713.17 & *       &  He I    & * & *\\ 
4740.17 & 4740.81 & [Ar III] & 1.30$\pm$0.097    & 1.33$\pm$0.10\\ 
4754.72 & 4756.13 & [Fe III] & 0.337$\pm$0.062   & 0.335$\pm$0.064\\
4861.33 & 4862.53 & H\,4     & 100.0$\pm$5.7    & 100.0$\pm$5.7\\
4881.11 & 4882.58 & [Fe III] & 0.703$\pm$0.072  & 0.700$\pm$0.071\\
4906.83 & 4907.85 &  O II    & 0.195$\pm$0.052  & 0.194$\pm$0.052\\
4921.93 & 4923.67 &  He I    & 0.240$\pm$0.057  & 0.862$\pm$0.080\\
4958.91 & 4960.14 & [O III]  & 358$\pm$20       & 351$\pm$21\\
5006.84 & 5008.17 & [O III]  & 1065$\pm$60      & 1035$\pm$63\\
5159.44 & 5159.12 & [Fe II]  & 0.219$\pm$0.030  & 0.207$\pm$0.030\\
5270.40 & 5270.83 & [Fe III] & 1.209$\pm$0.076  & 1.123$\pm$0.098\\ 
5411.52 & 5412.22 & He I     & 1.95$\pm$0.12    & 1.77$\pm$0.17\\
5517.66 & 5518.18 & [Cl III] & 0.789$\pm$0.056  & 0.71$\pm$0.08\\
5537.60 & 5538.29 & [Cl III] & 0.662$\pm$0.049  & 0.59$\pm$0.070\\
5679.56 & 5680.88 & N II     & 0.128$\pm$0.024  & 0.11$\pm$0.24\\
5754.60 & 5757.09 & [N II]   & 0.770$\pm$0.095  & 2.24$\pm$0.19\\
5875.66 & 5878.32 & He I     & 9.46$\pm$0.48    & 11.80$\pm$0.86\\
6300.34 & 6303.56 & [O I]    & 1.07$\pm$0.19    & 1.27$\pm$0.23\\
6312.10 & 6314.86 & [S III]  & 2.78$\pm$0.14    & 3.32$\pm$0.16\\
6363.78 & 6367.24 & [O I]    & 0.447$\pm$0.040  & 0.53$\pm$0.46\\
6548.10 & 6551.06 & [N II]   & 27.2$\pm$1.4     & 31.6$\pm$1.3\\
6562.77 & 6565.53 & H\,3     & 246$\pm$12       & 286$\pm$12\\
6583.50 & 6586.40 & [N II]   & 78$\pm$3.9       & 90.8$\pm$3.8\\
6678.16 & 6680.96 & He I     & 2.65$\pm$0.14    & 3.05$\pm$0.14\\
6716.44 & 6719.44 & [S II]   & 6.02$\pm$0.30    & 6.88$\pm$0.31\\
6730.82 & 6733.80 & [S II]   & 8.73$\pm$0.44    & 9.96$\pm$0.45\\
7065.25 & 7067.87 & He I     & 2.16$\pm$0.12    & 2.38$\pm$0.15\\
7135.80 & 7138.20 & [Ar III] & 11.41$\pm$0.57   & 12.50$\pm$0.80\\
7281.35 & 7284.05 &  He I    & 0.54$\pm$0.047   & 0.582$\pm$0.060\\
7319.99 & 7322.22 &  [O II]  & 3.56$\pm$0.18    & 3.83$\pm$0.29\\
7330.73 & 7333.22 &  [O II]  & 2.83$\pm$0.15    & 3.04$\pm$0.24\\
\\
\multicolumn{5}{l}{\footnotesize{c(H$\beta)\,=\,0.33\pm0.27$}}\\
\multicolumn{5}{l}{\footnotesize{$I_{\rm ob}\ (\mbox{H}\beta)\,=\,6.458\times10^{-12}\,\mbox{erg}\ \mbox{cm}^{-2}\ \mbox{s}^{-1}$}} \\
\multicolumn{5}{l}{\footnotesize{$I(\mbox{H}\beta)\,=\,1.381\times10^{-12}$}} \\
\multicolumn{5}{l}{\footnotesize{Extraction window = $2\arcsec\times14\arcsec$}} \\
\noalign{\vskip3pt} \noalign{\hrule} \noalign{\vskip3pt}
\multicolumn{5}{c}{\normalsize{NGC~3587}}\\
\noalign{\vskip3pt} \noalign{\hrule} \noalign{\vskip3pt}
3711.97 & 3710.54 & H\,15  & 117$\pm$25 & 151$\pm$48\\
3726.03 & 3727.29 & [O II] & 608$\pm$54 & 783$\pm$193\\
3728.82 & *       & [O II] & *  & *\\
3734.37 & 3734.05 & H\,13  & 41$\pm$13    & 53$\pm$21\\
3756.10 & 3758.73 & He I   & 28.7$\pm$7.7 & 37$\pm$13\\
3797.90 & 3798.16 & H\,10  & 29.8$\pm$8.8 & 38$\pm$14\\
3868.75 & 3868.09 & [Ne III] & 98$\pm$13 & 122$\pm$30\\
3967.46 & 3967.98 & [Ne III] & 39$\pm$11 & 48$\pm$16\\
3970.07 & *       & H\,7     & * & *\\
4083.90 & 4083.61 & O II     & 16.6$\pm$3.8 & 20.0$\pm$5.6\\
4085.11 & *       & O II     & *            &  *\\
\noalign{\vskip 3pt} \noalign{\hrule} \noalign{\vskip3pt}
\end{tabular}   
\end{minipage}  
\end{table}

\setcounter{table}{1}
\begin{table}
\begin{minipage}{75mm}
\centering \caption{{\it --continued}}
\begin{tabular}{c@{\hspace{2.8mm}}c@{\hspace{2.8mm}}c@{\hspace{1.8mm}}c@{\hspace{2.8mm}}c@{\hspace{2.8mm}}}
\noalign{\hrule} \noalign{\vskip2pt} \noalign{\hrule} 
$\lambda$(${\rm \AA}$)& $\lambda_{\rm ob}$(${\rm \AA}$) & Ion & $I_{\rm ob}(\lambda)$ & $I(\lambda)$\\
\noalign{\vskip3pt} \noalign{\hrule} \noalign{\vskip3pt}
\multicolumn{5}{c}{\normalsize{NGC~3587 (cont.)}}\\
\noalign{\vskip3pt} \noalign{\hrule} \noalign{\vskip3pt}
4101.74 & 4102.56 & H\,6     & 34.0$\pm$5.4 & 40.7$\pm$9.3\\
4340.63 & 4340.60 & H\,5     & 41.4$\pm$3.4 & 46.8$\pm$6.5\\
4363.21 & 4365.27 & [O III] & 7.99$\pm$0.55 & 8.45$\pm$0.96\\ 
4471.50 & 4473.72 & He I    & 4.91$\pm$0.41 & 5.12$\pm$0.55\\ 
4514.90 & 4515.34 & [Fe II]& 0.69$\pm$0.17 & 0.71$\pm$0.18\\ 
4514.86 & *       & N III  &    *          &   *\\ 
4641.81 & 4644.15 & O II   & 0.65$\pm$0.18 & 0.67$\pm$0.19\\ 
4641.84 & *       & N III  &    *        &   *\\ 
4643.08 & *       & N II   &    *        &   *\\ 
4685.68 & 4687.92 & He II  & 16.4$\pm$1.0   & 16.7$\pm$1.2\\ 
4701.62 & 4702.56 & [Fe III]&0.14$\pm$0.16  & 0.14$\pm$0.16\\
4711.37 & 4713.86 & [Ar IV]& 1.13$\pm$0.22  & 1.15$\pm$0.23\\ 
4713.17 & *       & He I   &		* & 	*\\
4740.17 & 4743.35 & [Ar IV] & 0.34$\pm$0.18 & 0.35$\pm$0.18\\ 
4861.33 & 4863.37 & H\,4    & 100.0$\pm$5.8 & 100.0$\pm$5.8\\ 
4921.93 & 4923.76 & He I  &  1.07$\pm$0.15  & 1.06$\pm$0.15\\ 
4958.91 & 4960.98 & [O III] & 299$\pm$17   & 296$\pm$18\\ 
5006.84 & 5008.87 & [O III] & 850$\pm$49   & 838$\pm$52\\ 
5411.52 & 5413.44 & He II   & 1.14$\pm$0.16 & 1.09$\pm$0.17\\
5517.66 & 5515.71 & [Cl III]& 0.54$\pm$0.10 & 0.72$\pm$0.15\\
5537.60 & 5536.92 & [Cl III]& 0.62$\pm$0.12 & 0.82$\pm$0.17\\
5754.60 & 5753.06 & [N II]  & 1.65$\pm$0.15 & 2.17$\pm$0.24\\
5875.66 & 5874.38 & He I    & 9.26$\pm$0.51  & 12.09$\pm$0.90\\
6233.80 & 6232.72 & He II   & 0.90$\pm$0.13 & 1.16$\pm$0.17\\
6300.34 & 6298.11 & [O I]   & 5.91$\pm$0.35 & 7.56$\pm$0.43\\
6310.80 & 6311.24 & He II   & 1.19$\pm$0.13 & 1.52$\pm$0.17\\
6312.10 & *       & [S III] &    *          &   *\\ 
6363.78 & 6361.78 & [O I]   & 1.92$\pm$0.16 & 2.45$\pm$0.20\\
6548.10 & 6546.50 & [N II]  & 32.4$\pm$1.7  & 40.9$\pm$1.8\\
6562.77 & 6561.26 &  H\,3   & 226.5$\pm$11.6& 286.3$\pm$12.3\\
6583.50 & 6581.87 & [N II]  & 94.6$\pm$4.9  & 119.2$\pm$5.1\\
6678.16 & 6676.58 & He I    & 2.61$\pm$0.16 & 3.27$\pm$0.18\\
6716.44 & 6714.74 & [S II]  & 18.01$\pm$0.93& 22.60$\pm$1.02\\
6730.82 & 6729.11 & [S II]  & 13.05$\pm$0.68& 16.36$\pm$0.75\\
7065.25 & 7063.38 & He I    & 1.53$\pm$0.14 & 1.88$\pm$0.18\\
7135.80 & 7133.55 & [Ar III]& 10.8$\pm$0.61 & 13.3$\pm$0.89\\
\\
\multicolumn{5}{l}{\footnotesize{c(H$\beta)\,=\,0.16\pm0.26$}}\\
\multicolumn{5}{l}{\footnotesize{$I_{\rm ob}\ (\mbox{H}\beta)\,=\,3.629\times10^{-13}\,\mbox{erg}\ \mbox{cm}^{-2}\ \mbox{s}^{-1} $}} \\
\multicolumn{5}{l}{\footnotesize{$I(\mbox{H}\beta)\,=\,5.246\times10^{-13}\mbox{erg}\ \mbox{cm}^{-2}\ \mbox{s}^{-1}$}} \\
\multicolumn{5}{l}{\footnotesize{Extraction window = $2\arcsec\times42\arcsec$}} \\
\noalign{\vskip3pt} \noalign{\hrule} \noalign{\vskip3pt}
\multicolumn{5}{c}{\normalsize{NGC~6210}}\\
\noalign{\vskip3pt} \noalign{\hrule} \noalign{\vskip3pt}
3679.36 & 3678.99 & H 21 & 0.82$\pm$0.55  & 0.94$\pm$0.65\\ 
3682.81 &    *    & H 20 &  * & * \\
3703.86 & 3703.35 & H 16 & 3.19$\pm$0.20  &  3.68$\pm$0.50\\
3705.02 &   *    & He I &      *         &  *\\
3711.97 & 3710.81 & H 15 & 1.26$\pm$0.11  & 1.45$\pm$0.22\\ 
3726.03 & 3726.58 & [O II] & 34.7$\pm$2.0 & 39.9$\pm$5.3\\ 
3728.82 &   *    & [O II] & &\\
3734.37 & 3734.36 & H 13 & 1.65$\pm$0.12  & 1.90$\pm$0.26\\
3750.15 & 3749.37 & H 12 & 2.73$\pm$0.19  & 3.13$\pm$0.42\\ 
3756.10 & 3756.41 & He I & 1.17$\pm$0.11  & 1.35$\pm$0.20\\
3757.21 &    *    & O III & * & *\\
3770.63 & 3770.16 & H 11 & 3.84$\pm$0.25  & 4.41$\pm$0.58\\ 
3797.90 & 3797.04 & H 10 & 4.26$\pm$0.26  & 4.88$\pm$0.63\\
3819.62 & 3818.55 & He I & 1.10$\pm$0.11  & 1.26$\pm$0.19\\ 
3835.39 & 3834.65 & H 9  & 6.78$\pm$0.40  & 7.73$\pm$0.97\\
3856.02  & 3857.28 & Si II & 0.40$\pm$0.08  & 0.46$\pm$0.10\\ 
3856.13  & *       & O II  & *              & *\\
3868.75  & 3868.05 & [Ne III] & 75.9$\pm$4.3 & 86$\pm$10\\ 
3888.65  & 3888.18 & H 8      & 17.8$\pm$1.0 & 20.2$\pm$2.4\\
3889.05  & *       & He I     & *            &\\
3907.46  & 3908.76 & O II     & 0.145$\pm$0.058  & 0.164$\pm$0.068 \\ 
3918.98  & 3917.87 & C II     & 0.252$\pm$0.072  & 0.285$\pm$0.087\\
3920.68  &  *      & C II     & * & *\\
3967.46  & 3967.66 & [Ne III] & 39.8$\pm$2.3     & 44.7$\pm$5.1\\  
3970.07  & *       &  H 7     &   * & *\\  
4009.26  & 4008.78 & He I     & 0.701$\pm$0.081  & 0.78$\pm$0.12\\  
4026.21  & 4025.47 & He I     & 2.47$\pm$0.16  & 2.75$\pm$0.31 \\ 
4068.60  & 4068.56 & [S II]   & 1.54$\pm$0.11  & 1.71$\pm$0.19 \\
4069.89  & *       &  O II    &   * & *\\  
\noalign{\vskip 3pt} \noalign{\hrule} \noalign{\vskip3pt}
\end{tabular}   
\end{minipage}  
\end{table}

\setcounter{table}{1}
\begin{table}
\begin{minipage}{75mm}
\centering \caption{{\it --continued}}
\begin{tabular}{c@{\hspace{2.8mm}}c@{\hspace{2.8mm}}c@{\hspace{1.8mm}}c@{\hspace{2.8mm}}c@{\hspace{2.8mm}}}
\noalign{\hrule} \noalign{\vskip2pt} \noalign{\hrule} 
$\lambda$(${\rm \AA}$)& $\lambda_{\rm ob}$(${\rm \AA}$) & Ion & $I_{\rm ob}(\lambda)$ & $I(\lambda)$\\
\noalign{\vskip3pt} \noalign{\hrule} \noalign{\vskip3pt}
\multicolumn{5}{c}{\normalsize{NGC~6210 (cont.)}}\\
\noalign{\vskip3pt} \noalign{\hrule} \noalign{\vskip3pt}
4072.16  & 4075.60 &  O II    &  0.681$\pm$0.076  & 0.76$\pm$0.11\\ 
4075.86  & *       & O II     & * & *\\
4076.35  & *       &  [S II]  &   * & *\\
4083.90  & 4083.58 & O II     & 0.066$\pm$0.038  & 0.073$\pm$0.043\\  
4085.11  & *       & O II     &   * & *\\ 
4087.15  & 4088.47 & O II     & 0.194$\pm$0.048  & 0.215$\pm$0.056\\ 
4089.29  & *       & O II    &   * & *\\
4092.93  & *       & O II    &   * & *\\ 
4101.74  & 4101.01 & H 6      & 25.1$\pm$1.4    &  27.8$\pm$2.8 \\ 
4120.84  & 4119.33 & He I      & 0.275$\pm$0.061 &  0.304$\pm$0.072\\
4143.76  & 4144.06 & He I      & 0.529$\pm$0.063 &  0.582$\pm$0.084\\
4257.80  & 4259.57 & Ne II     & 0.154$\pm$0.058 &  0.167$\pm$0.063\\ 
4267.15  & 4265.90 & C II      & 0.610$\pm$0.071 &  0.659$\pm$0.089\\
4275.55  & 4275.35 & O II & 0.193$\pm$0.048 &  0.208$\pm$0.054\\
4276.75  &  *  & O II      &   * & *\\
4283.73  & 4285.33 & O II     & 0.110$\pm$0.026 &  0.118$\pm$0.030\\
4285.69  & *   & O II      &   * & *\\
4315.40  & 4316.52   & O II      & 0.283$\pm$0.038 &  0.304$\pm$0.045\\
4315.83  & *   & O II      &   * & *\\
4317.14  & *   & O II      &   * & *\\
4340.47  & 4339.84 & H\,5       & 43.7$\pm$2.5    &  46.8$\pm$3.8\\ 
4363.21  & 4362.56 & [O III]   & 5.95$\pm$0.34   &  6.35$\pm$0.50\\ 
4379.11  & 4377.94 & N III     & 0.094$\pm$0.028 &  0.10$\pm$0.030\\ 
4379.55  & *       & Ne II     &   * & *\\
4387.93  & 4387.65 & He I      & 0.646$\pm$0.050 &  0.688$\pm$0.064\\ 
4391.94  &  *      & Ne II     & * & * \\ 
4409.30  & 4407.75 & Ne II     & 0.100$\pm$0.029 &  0.106$\pm$0.031\\
4413.11  & 4412.70  & Ne II     & 0.077$\pm$0.029 &  0.081$\pm$0.031\\
4413.22  & * & Ne II     &   * & *\\
4413.78  &  *      & [Fe II]   & * & *\\
4414.90  & * & O II      &   * & *\\
4428.54  & 4427.57 & Ne II     & 0.158$\pm$0.033 &  0.168$\pm$0.035\\
4430.94  & * & Ne II     &   * & *\\
4457.05  & 4455.12 & Ne II     & 0.076$\pm$0.033 &  0.080$\pm$0.035\\ 
4457.24  & * & Ne II     &   * & *\\
4465.41  & 4464.00 & O II      & 0.065$\pm$0.030 &  0.068$\pm$0.031\\ 
4471.50  & 4470.81 & He I      & 5.22$\pm$0.30   &  5.49$\pm$0.39\\ 
4571.10  & 4571.64 & Mg I]     & 0.188$\pm$0.038 &  0.195$\pm$0.040\\ 
4630.54  & 4634.30 & N II      & 0.466$\pm$0.039 &  0.480$\pm$0.042\\ 
4634.14  & * & N III     &   * & *\\
4638.86  & 4640.29 & O II      & 1.104$\pm$0.072 &  1.135$\pm$0.078\\
4640.64  & * & N III     &   * & *\\
4641.81  & * & O II & * & * \\
4649.13  & 4648.54 & O II     & 0.751$\pm$0.058 &  0.771$\pm$0.062\\ 
4650.84  & * & O II     &   * & *\\
4658.10  & 4658.80 & [Fe III] & 0.397$\pm$0.045 &  0.407$\pm$0.047\\
4661.63  & * & O II     &   * & *\\
4673.73  & 4673.26 & O II     & 0.174$\pm$0.036 &  0.178$\pm$0.038\\ 
4676.21  & * & O II     &   * & *\\
4685.68  & 4684.94 & He II    & 2.27$\pm$0.14   &   2.32$\pm$0.14\\ 
4696.35  & 4695.34 & O II     & 0.042$\pm$0.022 &  0.043$\pm$0.022\\
4699.22  & * & O II     &   * & *\\
4701.59  & * & [Fe III] & * & *\\
4711.37  & 4711.00 & [Ar IV]  & 2.09$\pm$0.13 & 2.13$\pm$0.13\\
4713.17  & * & He I     &   * & *\\
4740.17  & 4739.25 & [Ar IV]  & 1.563$\pm$0.098 &  1.59$\pm$0.10\\ 
4777.88  & 4779.30 & [Fe III] & 0.065$\pm$0.027 &  0.066$\pm$0.028\\ 
4783.34  & 4786.29 & O IV     & 0.073$\pm$0.029 &  0.074$\pm$0.030\\ 
4785.90  & * & C IV    &   * & *\\
4788.13  & * & N II     &   * & *\\
4861.33  & 4860.16  & H\,4     & 100.0$\pm$5.7     & 100.0$\pm$5.7\\ 
4881.11 &  4883.96  & [Fe III] & 0.0316$\pm$0.024  & 0.032$\pm$0.024\\ 
4921.93 &  4920.77  & He I     & 1.492$\pm$0.093   & 1.482$\pm$0.093\\ 
4924.53 &  *   & O II     &   * & *\\
4931.80 &  4931.37  & [O III]  & 0.558$\pm$0.049 & 0.554$\pm$0.049\\ 
4958.91 &  4957.53  & [O III]  & 346$\pm$20      & 342$\pm$20\\ 
5006.84 &  5005.38  & [O III]  & 985$\pm$56      & 968$\pm$57\\ 
5047.74 &  5046.44  &  He I   & 0.154$\pm$0.015  & 0.151$\pm$0.015\\
5191.82 &  5189.29  & [Ar III] & 0.054$\pm$0.012 & 0.053$\pm$0.012\\
5197.90 &  5197.10  & [N I]    & 0.127$\pm$0.014 & 0.123$\pm$0.014\\
5200.26 &       *   &  [N I]   & * & *\\
5342.38 &  5341.23  & C II     & 0.016$\pm$0.010 & 0.0157$\pm$0.0099\\
5411.52 &  5410.37  & He II    & 0.135$\pm$0.015 & 0.127$\pm$0.016\\
\noalign{\vskip 3pt} \noalign{\hrule} \noalign{\vskip3pt}
\end{tabular}   
\end{minipage}  
\end{table}

\setcounter{table}{1}
\begin{table}
\begin{minipage}{75mm}
\centering \caption{{\it --continued}}
\begin{tabular}{c@{\hspace{2.8mm}}c@{\hspace{2.8mm}}c@{\hspace{1.8mm}}c@{\hspace{2.8mm}}c@{\hspace{2.8mm}}}
\noalign{\hrule} \noalign{\vskip2pt} \noalign{\hrule} 
$\lambda$(${\rm \AA}$)& $\lambda_{\rm ob}$(${\rm \AA}$) & Ion & $I_{\rm ob}(\lambda)$ & $I(\lambda)$\\
\noalign{\vskip3pt} \noalign{\hrule} \noalign{\vskip3pt}
\multicolumn{5}{c}{\normalsize{NGC~6210 (cont.)}}\\
\noalign{\vskip3pt} \noalign{\hrule} \noalign{\vskip3pt}
5517.66 &  5515.53  & [Cl III] & 0.346$\pm$0.024 & 0.325$\pm$0.029\\
5537.60 &  5535.62  & [Cl III] & 0.415$\pm$0.028 & 0.389$\pm$0.034\\
5754.60 &  5753.91  & [N II]   & 0.527$\pm$0.035 & 0.457$\pm$0.034\\
5801.51 &  5800.08  & C IV     & 0.111$\pm$0.019 & 0.096$\pm$0.017\\
5812.14 &  5810.49  & C IV     & 0.110$\pm$0.019 & 0.095$\pm$0.017\\
5875.66 &  5874.48  & He I     & 18.20$\pm$0.91  & 15.67$\pm$0.86\\
5931.78 &  5931.30  & N II     & 0.0581$\pm$0.0097 & 0.0498$\pm$0.0084\\
5941.65 &  5942.91  & N II     & 0.036$\pm$0.011   &  0.0308$\pm$0.0091\\
6036.70 &  6036.99  & He II    & 0.0010$\pm$0.0085 &  0.0087$\pm$0.0072\\

6101.83 &  6100.16  & [K IV]   & 0.101$\pm$0.013   &  0.086$\pm$0.011\\
6157.42 &  6158.36  & Ni II    & 0.0214$\pm$0.0093 &  0.0181$\pm$0.0078\\
6157.60 & *         & [Mn V]   & * & *\\
6300.30 &  6299.32  & [O I]    & 2.66$\pm$0.14     &  2.23$\pm$0.10\\
6312.10 &  6310.85  & [S III]  & 1.322$\pm$0.069   &  1.108$\pm$0.052\\
6310.80 &  *  & He II    &  *                &       *\\ 
6363.78 &  6362.80  & [O I]    & 0.888$\pm$0.048   &  0.742$\pm$0.036\\
6461.95 &  6460.37  & C II     & 0.081$\pm$0.014   &  0.067$\pm$0.012\\
6548.10 &  6546.89  & [N II]   & 8.75$\pm$0.44     &  7.24$\pm$0.31\\
6562.77 &  6561.29  & H\,3     & 345$\pm$17        &  284.9$\pm$12.1\\
6583.50 &  6582.05  & [N II]   & 23.7$\pm$1.2      &  19.57$\pm$0.83\\
6678.16 &  6676.48  &  He I    & 5.18$\pm$0.26     &  4.25$\pm$0.19\\
6716.44 &  6714.88  & [S II]   & 2.80$\pm$0.15     &  2.30$\pm$0.11\\
6730.82 &  6729.20  & [S II]   & 4.64$\pm$0.24     &  3.80$\pm$0.17\\
7065.25 &  7062.96  &  He I    & 6.19$\pm$0.31     &  4.97$\pm$0.25\\
7135.80 &  7133.30  & [Ar III] & 13.01$\pm$0.65    &  10.40$\pm$0.53\\
7281.35 &  7279.45  & He I     & 0.63$\pm$0.04     &  0.498$\pm$0.032\\
7319.99 &  7316.92  & [O II]   & 2.82$\pm$0.15     &  2.23$\pm$0.13\\
7329.67 &  7327.67  & [O II]   & 2.38$\pm$0.12     &  1.88$\pm$0.11\\
\\
\multicolumn{5}{l}{\footnotesize{c(H$\beta)\,=\,0.19\pm0.16$}}\\
\multicolumn{5}{l}{\footnotesize$I_{\rm o}\ (\mbox{H}\beta)\,=\,2.083\times10^{-14}\,\mbox{erg}\ \mbox{cm}^{-2}\ \mbox{s}^{-1}$}\\
\multicolumn{5}{l}{\footnotesize$I(\mbox{H}\beta)\,=\,3.226\times10^{-13} \mbox{erg}\,\mbox{cm}^{-2}\ \mbox{s}^{-1}$}\\
\multicolumn{5}{l}{\footnotesize Extraction window = $2\arcsec\times16\arcsec$}\\
\noalign{\vskip 3pt} \noalign{\hrule} \noalign{\vskip3pt}
\end{tabular}   
\end{minipage}  
\end{table}

The errors in the line intensities were obtained by adding 
quadratically: (1) the error due to the flux calibration (4\% in the blue
range and 3\% in the red range) derived from the standard 
deviation in the calibration curves of the standard stars;
(2) the statistical errors associated with the measurement of the line
intensities, which have been calculated using 
$\sigma_{\rm l}=\sigma_{\rm c}\sqrt{N + EW/\Delta}$ \citep{Perez},
where $\sigma_{\rm l}$ is the error in the observed line intensity, 
$\sigma_{\rm c}$ represents the standard deviation in a box near 
the measured emission line and stands for the error in the continuum 
placement, $N$ is the number of pixels used in the measurement of the 
line intensity, EW is the line equivalent width, and $\Delta$ is the
wavelength  dispersion in \AA\ pixel$^{-1}$; and (3) the error associated 
with the extinction correction.

\begin{figure*}
\centering
\begin{tabular}{cc}
\subfigure{
\includegraphics[width=9cm,trim = 50 0 10 0,clip =yes]{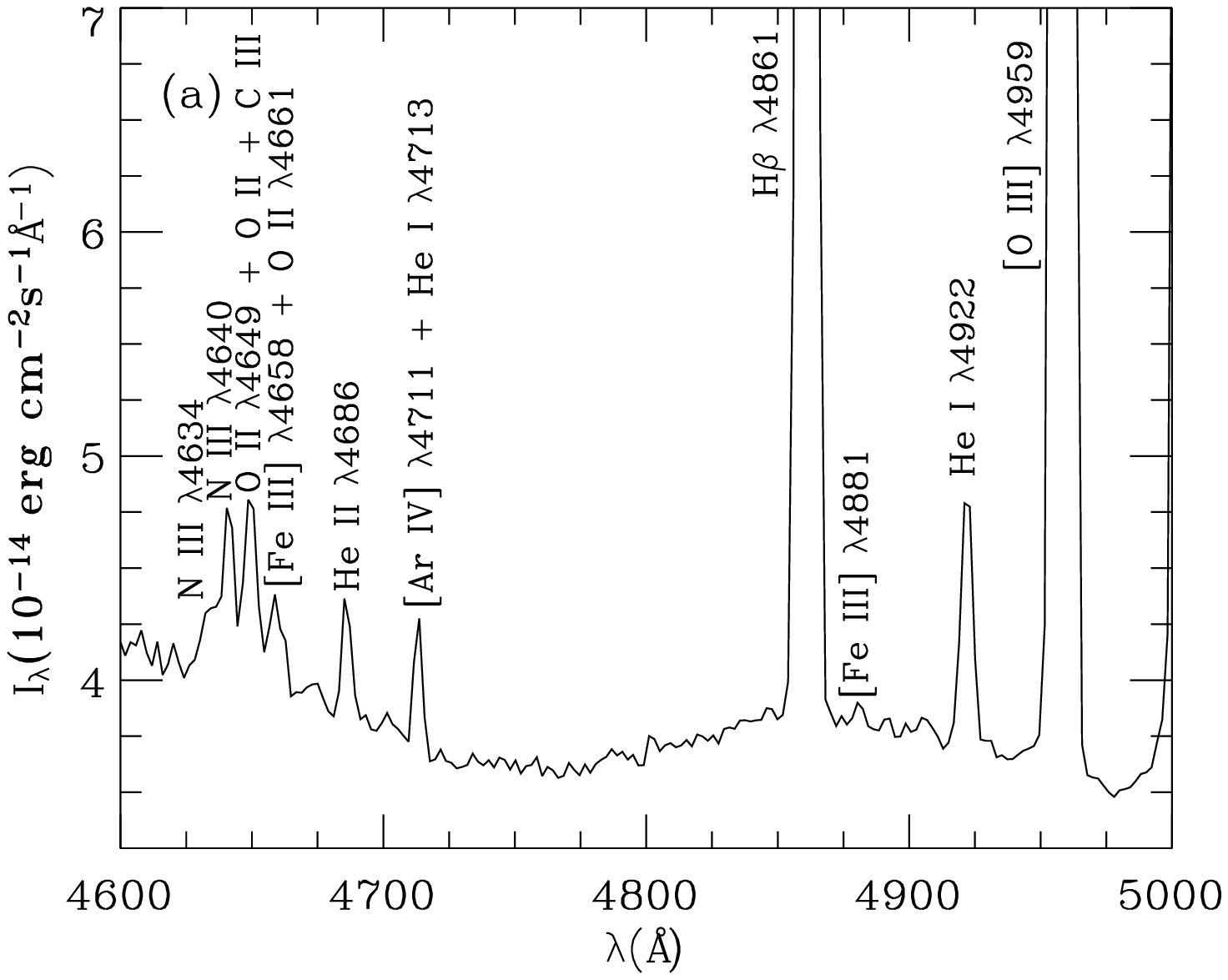}}
\subfigure{
\includegraphics[width=9cm,trim = 50 0 10 0,clip =yes]{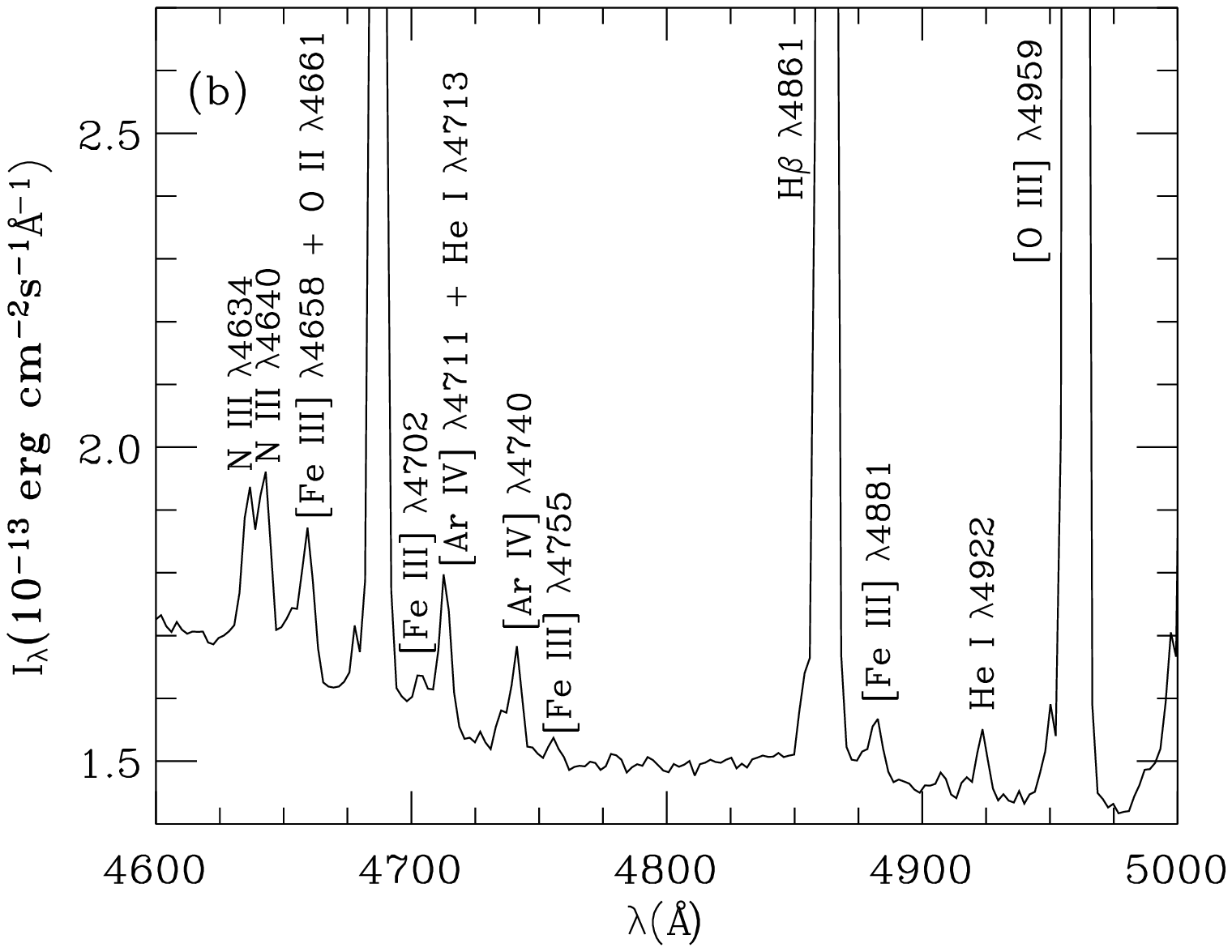}}\\
\subfigure{
\includegraphics[width=9cm,trim = 50 20 10 30,clip =yes]{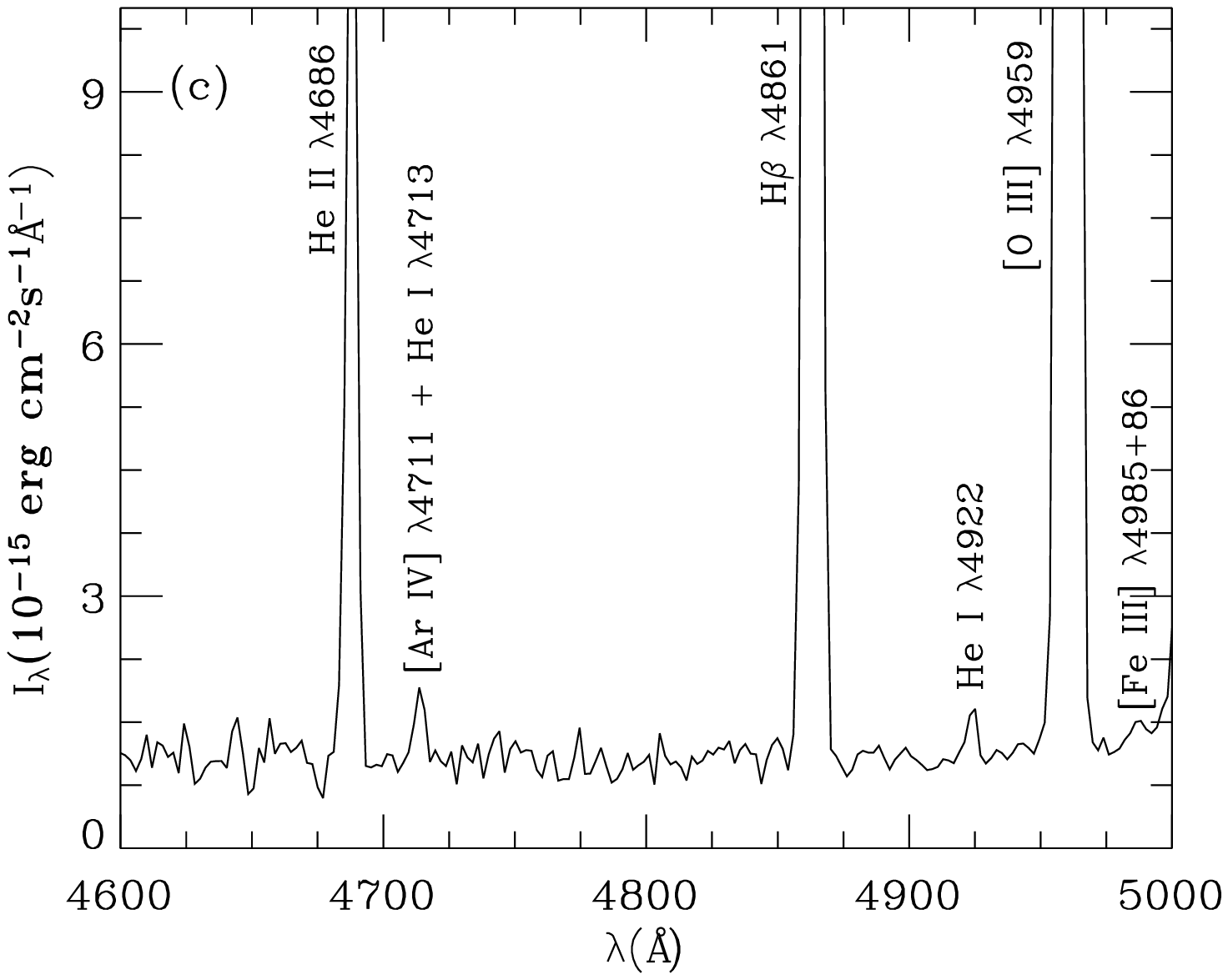}}
\subfigure{
\includegraphics[width=9cm,trim = 50 20 10 30,clip =yes]{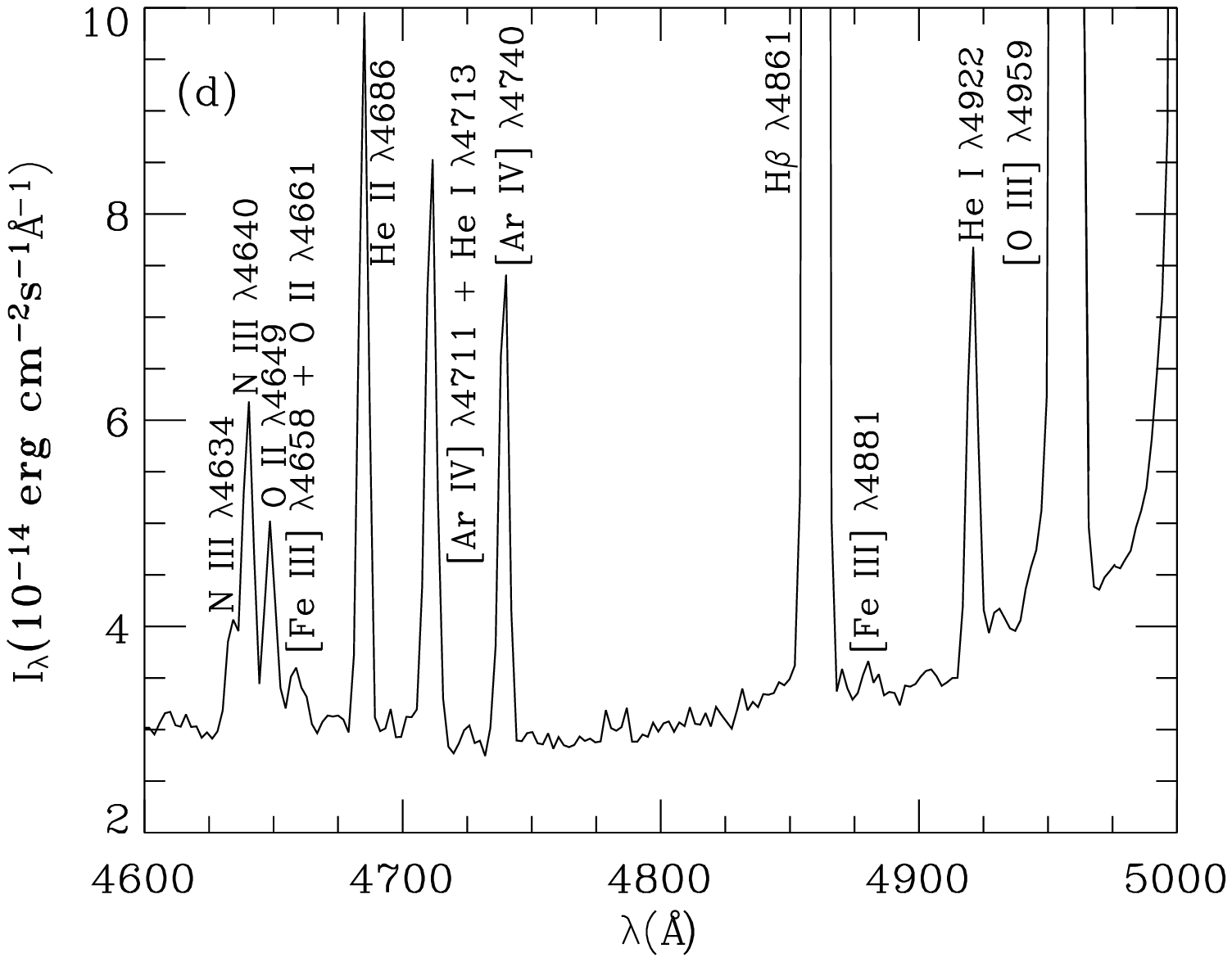}}
\end{tabular}
\caption{{\small Part of the blue spectra of (a) IC~4593, (b) NGC~2392, 
(c) NGC~3587, and (d) NGC~6210 showing the region where we find 
most of the [\ion{Fe}{3}] lines.}\label{lineas_Fe1}}
\end{figure*}

\begin{figure}
\subfigure{
\includegraphics[width=9cm,trim = 50 30 10 0,clip =yes]{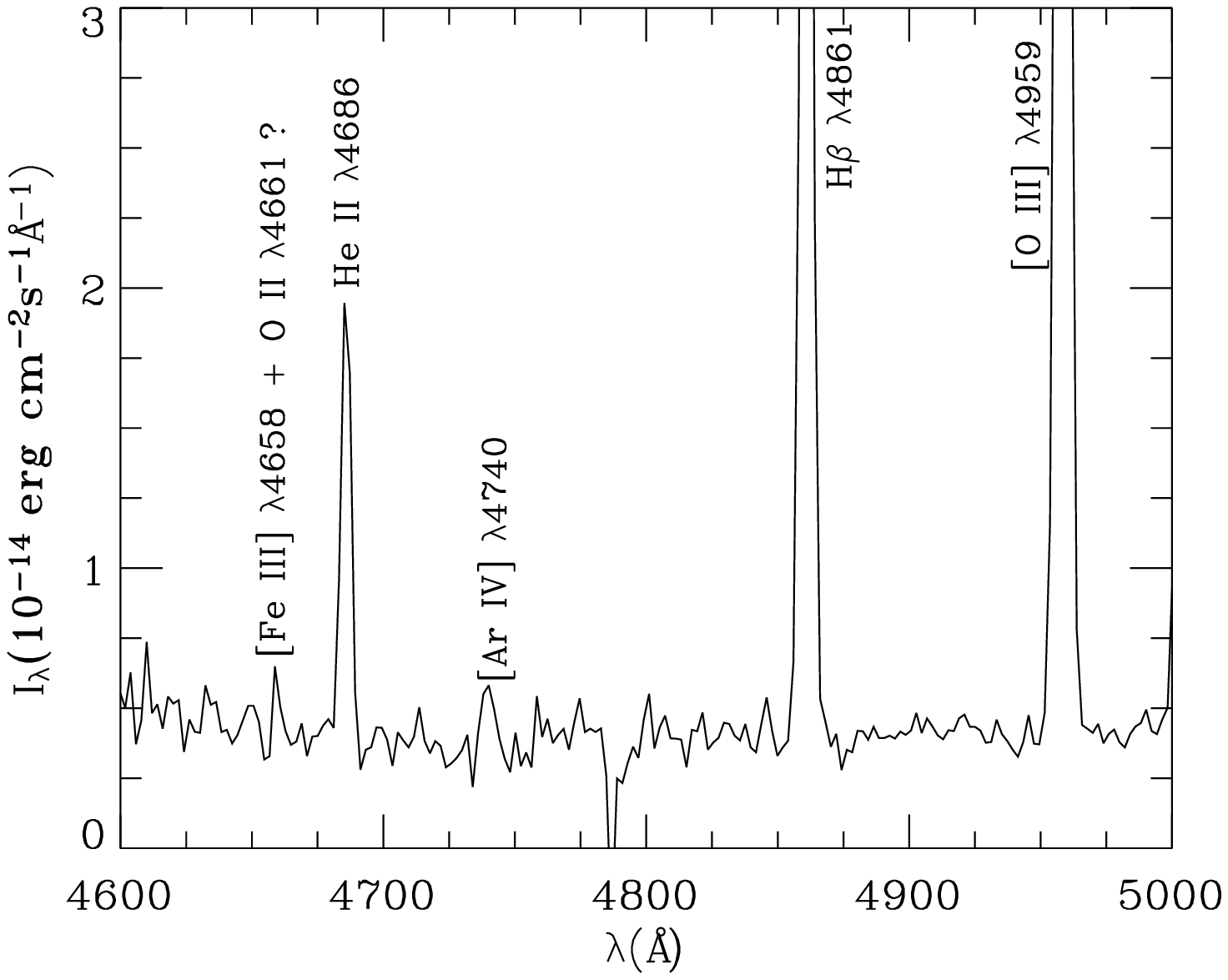}}\\
\caption{{\small Part of the blue spectra of JnEr~1 showing the feature used 
to derive upper limits to the Fe abundance.}\label{lineas_Fe2}}
\end{figure}

Figures~\ref{lineas_Fe1} and \ref{lineas_Fe2} show the
$\lambda\lambda$4600--5000 spectral region of the observed objects,
where most of the [\ion{Fe}{3}] lines are located. 

\section{Physical conditions and ionic abundances \label{condiciones}} 

We calculated physical conditions and ionic abundances for the 
five PNe we observed, and recalculated them from the line intensities
available in the literature for the rest of the PNe in our sample. 
We used the diagnostic line ratios [\ion{N}{2}] 
$\lambda5755/(\lambda6548+\lambda6584)$ and 
[\ion{O}{3}] $\lambda4363/(\lambda4959+\lambda5007)$ to derive the 
electron temperatures (\T) of the low and high-ionization regions, respectively. 
The adopted electron density for each PN is the weighted mean of 
the \n\ values obtained from the available line ratios of the three diagnostics
we used: [\ion{S}{2}] $\lambda6716/\lambda6731$, 
[\ion{Cl}{3}] $\lambda5518/\lambda5538$ and [\ion{Ar}{4}] 
$\lambda4711/\lambda4740$.
Table \ref{tabla_3} shows the values of \T\ and \n\ we derived for 
each PN of the sample, and the references for the line intensities.
The maximum differences between the physical conditions we obtained 
and the ones presented in the literature are of about 20\% for \T\ 
and 40\% for \n; these differences are mostly due to the use of 
different atomic data.

\begin{deluxetable*}{llccccccc}
\tabletypesize{\small}
\tablecaption{Physical Conditions\label{tabla_3}}
\tablewidth{0pt}
\tablehead{
\colhead{Object} & \colhead{PNG\tablenotemark{a}} & \colhead{\TN} & \colhead{\TO} & \colhead{\dens} & \colhead{\dencl} & \colhead{\denar} & \colhead{\n}(adopted) & \colhead{Ref.}\\
                 &                & \colhead{(K)} & \colhead{(K)} & \colhead{(\cm)} & \colhead{(\cm)} & \colhead{(\cm)} & \colhead{(\cm)} &}
\startdata
\objectname{Cn 3-1}& 038.2$+$12.0 & 	$7600\pm300$            & $7700^{+500}_{-300}$   & $9400^{+12300}_{-3600}$ & $12200^{+3200}_{-2300}$ & \nodata & $11900\pm1600$   & 1\\	
\objectname{Hu 1-1}& 119.6$-$06.7 & 	$11500\pm400$           & $12100\pm300$          & $1500^{+400}_{-300}$ & $<100$ & $2600^{+1100}_{-900}$ & $1600\pm300$      & 1\\
\objectname{IC 418}& 215.2$-$24.2 & 	$9100^{+600}_{-500}$    & $8800\pm200$           & $29200_{-19100}$ & $13400^{+3000}_{-2200}$ & $7000^{+8200}_{-3900}$  &$12400\pm2400$   & 2\\
\objectname{IC 1747}& 130.2$+$01.3 & 	$12300^{+700}_{-600}$   & $10700^{+300}_{-200}$  & $5000^{+2300}_{-1300}$ & \nodata & $3600^{+1100}_{-900}$ & $3900\pm900$      & 1\\	 
\objectname{IC 3568}& 123.6$+$34.5 & 	$18800^{+4500}_{-2600}$ & $11400\pm300$          & $2000^{+1500}_{-800}$ & \nodata & $2500^{+1000}_{-900}$ & $2300\pm700$      & 3\\	
\objectname{IC 4191}& 304.5$-$04.8 & 	$11100^{+600}_{-500}$   & $9900\pm200$           & $8300^{+6400}_{-2700}$  & $13600^{+2400}_{-1900}$  & $17200^{+2900}_{-2300}$ & $14900\pm1400$  & 4\\
\objectname{IC 4406}& 319.6$+$15.7 & 	$10300\pm300$           & $10000\pm200$          & $1000^{+300}_{-200}$ & $4500^{+900}_{-800}$ & $1500^{+900}_{-700}$ & $1300\pm200$       & 4\\
\objectname{IC 4593}& 025.3$+$40.8 & 	9700$^{+1300}_{-900}$  & 8500$^{+300}_{-200}$   & $2000^{+700}_{-500}$ & $800^{+1600}$ & \nodata & 1900$\pm500$	   & 5\\
\objectname{IC 4846}& 027.6$-$09.6 & 	$11900^{+3100}_{-700}$  & $10500^{+500}_{-400}$  & $6900^{+18900}_{-3200}$ & $12600^{+63100}_{-7400}$ & $10700^{+4400}_{-3000}$ & $10400\pm3300$   & 6\\  
\objectname{IC 5217}&	100.6$-$05.4 & 	$13600^{+6000}_{-2500}$ & $10700^{+500}_{-400}$  & $4800^{+3900}_{-1700}$ & $4300^{+3500}_{-2000}$ & $6400^{+4500}_{-2800}$ & $5000\pm1700$	   & 7\\ 
\objectname{JnEr 1}&	164.8$+$31.1 & 	10300$^{+1000}_{-800}$  & 11700$^{+800}_{-600}$  & $200^{+500}$ & \nodata & \nodata &  200$^{+500}_{-180}$& 5\\
\objectname{M 1-73}&	051.9$-$03.8 & 	$8500^{+600}_{-500}$    & $7200\pm100$          & $5900^{+3800}_{-1800}$ & \nodata & \nodata & $5900\pm2800$      & 1\\
\objectname{MyCn 18}&	307.5$-$04.9 & 	$9700^{+500}_{-400}$    & $7300\pm100$           & $5500^{+3000}_{-1500}$ & $12800^{+2300}_{-1800}$ & \nodata & $9500\pm1500$      & 4\\
\objectname{NGC 40}&	120.0$+$09.8 & 	$8600^{+300}_{-200}$    & $10600^{+300}_{-200}$  & $1800^{+600}_{-400}$ & $1100^{+500}_{-400}$ & \nodata & $1400\pm300$	   & 3\\
\objectname{NGC 2392}&	197.8$+$17.3 & 	$12700^{+2200}_{-1400}$ & $14500^{+1100}_{-900}$ & $2600^{+22000}_{-1600}$ & $2100^{+2000}_{-1300}$ & $1200^{+1800}$ & $1700\pm1200$	   & 5\\	
\objectname{NGC 3132}&	272.1$+$12.3 & 	$9700^{+300}_{-200}$      & $9500\pm200$         & $600^{+200}_{-100}$ & $800^{+500}_{-400}$ & $700^{+800}_{-700}$ & $600\pm100$	   & 4\\ 
\objectname{NGC 3242}&	261.0$+$32.0 & 	$12100^{+1700}_{-1100}$ & $11800\pm300$          & $2300^{+700}_{-500}$ & $1300^{+600}_{-500}$ & $3300^{+1100}_{-900}$ & $2000\pm400$	   & 4\\ 	
\objectname{NGC 3587}&	148.4$+$57.0 & 	9800$^{+900}_{-700}$    & $11600\pm500$          & $<100$ & \nodata & \nodata &  100$^{+400}_{-80}$& 5\\
\objectname{NGC 5882}&	327.8$+$10.0 & 	$10500^{+500}_{-400}$   & $9400\pm200$           & $4900^{+2400}_{-1300}$ & $4400^{+1000}_{-800}$ & $6400^{+1400}_{-1200}$ & $5000\pm700$  & 4\\ 
\objectname{NGC 6153}&	341.8$+$05.4 & 	$10300\pm400$           & $9100\pm200$           & $4200^{+1800}_{-1100}$ & $5000^{+1000}_{-900}$ & $3000^{+1000}_{-800}$ & $4000\pm600$ & 8\\
\objectname{NGC 6210}&	043.1$+$37.7 & 	$11600^{+700}_{-600}$   & 10000$\pm300$          & $4300^{+1600}_{-1000}$ & $4400^{+1800}_{-1300}$ & $7200^{+2900}_{-2100}$ & 4700$\pm$900   & 5\\
\objectname{NGC 6210}&	043.1$+$37.7 & 	11100$^{+500}_{-400}$   & 9600$\pm200$           & $4200^{+1800}_{-1100}$ & $4100^{+900}_{-800}$ & $9000^{+1700}_{-1400}$ & 5000$\pm$700	   & 3\\
\objectname{NGC 6543}&	096.4$+$29.9 & 	$10000\pm400$           & $7900\pm100$           & $6900^{+4900}_{-2200}$ & $6400^{+1200}_{-1000}$ & $4400^{+1100}_{-900}$  & $5400\pm700$	   & 9\\  
\objectname{NGC 6572}&	034.6$+$11.8 & 	$11200^{+800}_{-600}$   & $10300^{+300}_{-200}$  & $25200_{-13600}$ & $20000^{+3800}_{-2900}$ & $23200^{+3800}_{-3100}$ & $21600\pm2400$   & 3\\ 
\objectname{NGC 6720}&	063.1$+$13.9 & 	$10600\pm300$           & $10600^{+300}_{-200}$  & $500^{+200}_{-100}$ & $500^{+500}_{-400}$ & $1400^{+900}_{-700}$ & $500\pm100$	   & 3\\   
\objectname{NGC 6803}&	046.4$-$04.1 & 	$10200^{+500}_{-400}$   & $9600\pm200$            & $8300^{+6800}_{-2700}$ & $11600^{+2100}_{-1700}$ & $12900^{+2200}_{-1800}$ & $11900\pm1300$   & 1\\
\objectname{NGC 6826}&	083.5$+$12.7 & 	$10600^{+600}_{-500}$   & $9300\pm200$           & $1900^{+600}_{-400}$ & $1600^{+600}_{-500}$ & $3100^{+1000}_{-900}$ & $1900\pm300$	  & 3\\   
\objectname{NGC 6884}&	082.1$+$07.0 & 	$11500^{+600}_{-500}$   & $11000\pm300$          & $8200^{+6400}_{-2600}$ & $6900^{+1300}_{-1100}$ & $14500^{+2500}_{-2000}$ & $8600\pm1000$      & 3\\
\objectname{NGC 7026}&	089.0$+$00.3 & 	$9300\pm400$            & $9200\pm200$           & \nodata & $15600^{+4300}_{-3000}$ & $7700^{+1500}_{-1300}$ & $9100\pm1100$      & 1
\enddata
\tablerefs{{\it Line intensities from:} (1) \citet{Wesson_05},
(2) \citet{Sharpee_03}, (3) \citet{Liu_04b}, (4) \citet{Tsamis_03}, (5) this work, 
(6) \citet{Hyung_01b}, (7) \citet{Hyung_01a}, (8) \citet{Liu_00}, (9) \citet{Wesson_04}.}\\
\tablenotetext{a}{PNG identifications from \citet{Acker_92}.}
\end{deluxetable*}

The [\ion{N}{2}] $\lambda5755$ line can be affected by recombination 
excitation, and \citet{Liu_00} derived an expression that can be used 
to correct for this effect the value of \TN.
We checked and found that the effect of this correction in the total abundances
of O and Fe is not important for the PNe in our sample: if we had considered
the contribution of recombination excitation, the derived total abundances
would be consistent within the errors with those presented here.
For this reason, and because the value of the correction is somewhat uncertain,
we did not take this effect into account. 

We used the values of \TN\ to derive the O$^{+}$ and Fe$^{++}$ 
abundances, and \TO\ for He$^{+}$, He$^{++}$, and O$^{++}$. We calculated 
the He$^{+}$ and  He$^{++}$ abundances in order to estimate the contribution 
to the total O abundance of ions of higher degree of ionization 
than O$^{++}$ (see more details in Section~5). 
The physical conditions and the $\mbox{O}^{+}/\mbox{H}^{+}$ and 
$\mbox{O}^{++}/\mbox{H}^{+}$ abundance ratios were calculated with the
{\sc IRAF} {\sc NEBULAR} package. 
However, we changed the atomic data for [\ion{Cl}{3}] and used the transition
probabilities of \citet{Mendoza_82} and the collision strengths of
\citet{Mendoza_83}, because these data lead to densities that are in better
agreement with the values implied by the other diagnostics.
To derive the $\mbox{He}^{+}$ abundance we used the calculations of
\citet{Benjamin_99} and the \ion{He}{1} $\lambda$6678 line, since it is the
brightest of the measured singlet lines and singlet lines are not 
affected by self-absorption effects. The 
$\mbox{He}^{++}$ abundance was calculated using the \ion{He}{2} 
$\lambda$4686 line and the emissivities of \citet{Storey_95}. We also
used the emissivities of \citet{Storey_95} for \ion{H}{1}. 
The errors in the
O$^{+}$ and O$^{++}$ abundances have been derived by adding quadratically 
the errors in the line intensity ratio used and the errors arising 
from the uncertainties in \T\ and \n. Columns 2, 3, 4, 
and 5 in Table~\ref{tabla_4} show our derived 
ionic abundances for He$^{+}$, He$^{++}$, O$^{+}$, and O$^{++}$.

\begin{deluxetable*}{lllllllclcll}
\tabletypesize{\small}
\tablecaption{Ionic and Total Abundances:  \{X$^{+i}$\} = 12 + $\log$ (X$^{+i}$/H$^{+}$), 
\{X\} = 12 + $\log$ (X/H)\label{tabla_4}}
\tablewidth{0pt}
\tablehead{
\colhead{Object\tablenotemark{a}} & \colhead{\{He$^{+}$\}} & \colhead{\{He$^{++}$\}} & \colhead{\{O$^{+}$\}} & \colhead{\{O$^{++}$\}} &  ICF & \colhead{\{O\}} & \colhead{\{Fe$^{+}$\}} & \colhead{\{Fe$^{++}$\}} & \colhead{N\tablenotemark{b}} 
& \colhead{\{Fe\}\tablenotemark{c}}  & \colhead{\{Fe\}\tablenotemark{d}}\\
\colhead{(1)} & \colhead{(2)} & \colhead{(3)} & \colhead{(4)} & \colhead{(5)} & \colhead{(6)} & \colhead{(7)} 
& \colhead{(8)} & \colhead{(9)} & \colhead{(10)} & \colhead{(11)} & \colhead{(12)}}
\startdata
Cn3-1	& 10.67 &  7.52 & $8.80\pm0.11$          & $7.27^{+0.08}_{-0.10}$  & 1.00  & $8.81\pm0.11$          & -- & $5.62^{+0.04}_{-0.05}$  & 3   & 5.71$^{+0.14}_{-0.16}$  & 5.63$^{+0.15}_{-0.17}$ \\	
Hu~1-1	& 10.93 & 10.18 & $7.99\pm0.06$          & $8.38^{+0.03}_{-0.04}$  & 1.11  & $8.57\pm0.03$          & -- & $4.25^{+0.16}_{-0.25}$  & 1   & $4.76^{+0.17}_{-0.27}$  & $4.65^{+0.17}_{-0.27}$ \\	
IC~418	& 10.97 &  --   & $8.47^{+0.13}_{-0.15}$ & $8.08\pm0.03$           & 1.00  & $8.62\pm0.10$          & 3.93: & $4.14^{+0.06}_{-0.08}$ & 10 & $4.27^{+0.16}_{-0.20}$ & $4.50^{+0.16}_{-0.20}$ \\
IC~1747	& 11.01 & 10.04 & 7.09$^{+0.09}_{-0.10}$ & $8.54\pm0.03$           & 1.07  & $8.58\pm0.03$          & --& $<4.34$                   & --  & $<5.67$                & $<5.04$ 			\\	
IC~3568	& 10.96 &  9.02 & $5.72^{+0.18}_{-0.09}$ & $8.37\pm0.04$           & 1.01  & $8.37\pm0.04$          & -- &$3.82^{+0.12}_{-0.16}$   & 2   & $6.21^{+0.20}_{-0.20}$ & 4.98$^{+0.16}_{-0.22}$ 	\\	
IC~4191 & 11.05 & 10.08 & $7.56^{+0.09}_{-0.10}$ & $8.70\pm0.03$           & 1.07  & $8.76\pm0.03$          & -- &$4.42^{+0.10}_{-0.12}$   & 1   & 5.48$^{+0.12}_{-0.16}$ & 5.00$^{+0.11}_{-0.15}$ 	\\
IC~4406 & 10.97 & 10.08 & $8.24\pm0.06$          & $8.57\pm0.03$           & 1.08  & $8.77\pm0.03$          & -- &$<4.55$                  & --  & $<5.02$	              & $<4.94$ 		\\
IC~4593 & 11.01 &  8.54 & 7.45$^{+0.20}_{-0.24}$ & $8.55\pm0.04$           & 1.00  & $8.58\pm0.04$          & -- &$5.36^{+0.09}_{-0.11}$   & 4   & 6.36$^{+0.20}_{-0.27}$ & 5.90$^{+0.14}_{-0.18}$ 	\\
IC~4846 & 10.97 & 8.69  & $7.07^{+0.14}_{-0.39}$ & $8.50^{+0.04}_{-0.05}$ & 1.00   & $8.51^{+0.04}_{-0.05}$ & -- &$4.55^{+0.10}_{-0.13}$   & 3   & 5.83$^{+0.16}_{-0.41}$ & 5.20$^{+0.14}_{-0.23}$ 	\\  
IC~5217 & 10.84 & 9.96  & $6.60^{+0.33}_{-0.47}$ & $8.63^{+0.04}_{-0.05}$ & 1.08   & $8.67^{+0.04}_{-0.05}$ & -- &$4.62^{+0.26}_{-0.45}$   & 1   & 6.49$^{+0.37}_{-0.98}$ & 5.56$^{+0.30}_{-0.58}$ 	\\ 
JnEr~1  & 11.32 & 10.24 & 8.42$^{+0.16}_{-0.17}$ & 7.98$^{+0.05}_{-0.06}$ & 1.06   & $8.58\pm0.12$          & -- &$<6.00$                  & --  & $<6.15$		      & $<6.16$ 		\\
M~1-73  & 11.03 & 9.00  & $8.27^{+0.16}_{-0.20}$ & $8.54^{+0.03}_{-0.04}$ & 1.01   & $8.73\pm0.06$          & -- &$5.45^{+0.08}_{-0.10}$   & 2   & 5.85$^{+0.17}_{-0.24}$ & 5.80$^{+0.13}_{-0.16}$  \\
MyCn~18 & 10.95 & 8.66  & $7.83^{+0.09}_{-0.11}$ & $8.52\pm0.02$          & 1.00   & $8.60\pm0.03$          & -- &$5.47\pm0.03$            & 6   & 6.14$^{+0.09}_{-0.11}$ & 5.88$^{+0.07}_{-0.08}$ \\
NGC~40 & 10.80 & 7.56  & $8.63^{+0.06}_{-0.08}$ & $7.07^{+0.03}_{-0.05}$ & 1.00   & $8.64^{+0.06}_{-0.08}$ & 4.86:  & $5.58\pm0.02$       & 7   & 5.67$^{+0.08}_{-0.11}$ & 5.67$^{+0.08}_{-0.12}$  \\
NGC~2392 & 10.90 & 10.46 & 7.41$^{+0.19}_{-0.26}$ & 8.07$^{+0.05}_{-0.06}$ & 1.23   & 8.25$\pm0.06$          & -- &$5.62^{+0.07}_{-0.08}$   & 5   & 6.35$^{+0.19}_{-0.27}$ & 6.11$^{+0.14}_{-0.19}$  \\	
NGC~3132 & 11.04 & 9.51  & $8.42^{+0.05}_{-0.06}$ & $8.52\pm0.03$          & 1.02   & $8.78\pm0.03$          & -- &$4.99^{+0.03}_{-0.04}$   & 6   & 5.30$^{+0.06}_{-0.08}$ & 5.35$^{+0.06}_{-0.08}$  \\
NGC~3242 & 10.90 & 10.34 & $6.53^{+0.16}_{-0.20}$ & $8.41\pm0.03$          & 1.18   & $8.49\pm0.03$          & -- &$3.84^{+0.12}_{-0.17}$   & 2   & 5.60$^{+0.18}_{-0.28}$ & 4.75$^{+0.15}_{-0.21}$ \\ 	
NGC~3587 & 10.94 & 10.15 & 8.09$^{+0.15}_{-0.16}$ & $8.27^{+0.04}_{-0.05}$ & 1.11   & $8.53\pm0.07$          & -- &$5.24^{+0.36}_{-0.27}$   & 1   & $5.62^{+0.37}_{-0.36}$ & $5.62^{+0.37}_{-0.32}$ \\
NGC~5882   & 11.02 & 9.35  & $6.92^{+0.08}_{-0.09}$ & $8.66\pm0.03$          & 1.01   & $8.67\pm0.03$          & -- &$4.73^{+0.04}_{-0.05}$   & 6   & 6.30$^{+0.09}_{-0.11}$ & 5.52$^{+0.08}_{-0.09}$ \\ 
NGC~6153   & 11.06 & 10.05 & $7.20^{+0.08}_{-0.08}$ & $8.62\pm0.03$          & 1.06   & $8.67\pm0.03$          & -- &$4.50^{+0.06}_{-0.08}$   & 3   & 5.80$^{+0.10}_{-0.12}$ & 5.18$^{+0.09}_{-0.11}$ \\
NGC~6210\tablenotemark{e}   & 10.99 & 9.09  & 7.16$^{+0.10}_{-0.11}$ & 8.54$^{+0.03}_{-0.04}$ & 1.01   & 8.56$\pm0.03$ & -- &$4.62^{+0.08}_{-0.10}$ & 2   & 5.87$^{+0.12}_{-0.15}$ & 5.26$^{+0.11}_{-0.14}$ \\
NGC~6210\tablenotemark{f}   & 11.01 & 9.28  & $7.21^{+0.07}_{-0.08}$ & 8.62$\pm0.04$          & 1.01   & 8.64$\pm0.04$ & -- &$4.60^{+0.04}_{-0.05}$ & 8   & 5.88$^{+0.08}_{-0.10}$ & 5.26$^{+0.09}_{-0.11}$ \\
NGC~6543   & 11.05 & --    & $7.24\pm0.08$          & $8.75\pm0.03$          & 1.00   & 8.76$\pm0.03$          & -- &$4.90\pm0.04$            & 7   & 6.26$^{+0.09}_{-0.10}$ & 5.59$^{+0.08}_{-0.09}$ \\  
NGC~6572   & 11.01 & 8.53  & $7.59^{+0.10}_{-0.13}$ & $8.58\pm0.04$          & 1.00   & 8.62$^{+0.03}_{-0.04}$ & -- &$4.55\pm0.06$            & 7   & 5.46$^{+0.11}_{-0.15}$ & 5.05$^{+0.09}_{-0.13}$ \\  
NGC~6720   & 10.97 & 10.25 & $8.21\pm0.06$           & $8.47^{+0.03}_{-0.05}$& 1.12   & $8.71^{+0.03}_{-0.04}$ & 4.28: &$4.70^{+0.04}_{-0.05}$ & 4  & 5.14$^{+0.07}_{-0.08}$ & 5.09$^{+0.07}_{-0.09}$ \\   
NGC~6803   & 11.04 & 9.56  & $7.51^{+0.08}_{-0.10}$  & $8.64\pm0.04$         & 1.02   & $8.68\pm0.04$          & -- &$4.90\pm0.04$            & 3   & 5.93$^{+0.09}_{-0.11}$ & 5.45$^{+0.08}_{-0.10}$ \\
NGC~6826   & 11.00 & 7.34  & $6.99^{+0.09}_{-0.11}$  & $8.52\pm0.04$         &  1.00  & $8.53\pm0.04$          & -- &$4.72\pm0.05$            & 5   & 6.10$^{+0.10}_{-0.12}$ & 5.42$^{+0.10}_{-0.12}$ \\   
NGC~6884   & 10.87 & 10.19 & $7.16^{+0.08}_{-0.09}$  & $8.55\pm0.04$         & 1.13   & $8.62\pm0.04$          & 4.04: &$4.74\pm0.04$         & 6   & 6.04$^{+0.09}_{-0.11}$ &5.44$^{+0.10}_{-0.11}$  \\
NGC~7026   & 11.04 & 10.12 & 7.86$\pm0.10$           & $8.62\pm0.04$         & 1.08   & $8.72\pm0.04$          & -- &$<4.73$                  & --  & $<5.49$		      & $<5.19$ 
\enddata
\tablenotetext{a}{Line intensities from the same references of Table~\ref{tabla_3}.}
\tablenotetext{b}{Number of [\ion{Fe}{3}] lines used in the Fe$^{++}$ abundance
determination.}
\tablenotetext{c}{Derived using equation~(\ref{eq1}).}
\tablenotetext{d}{Derived using equations~(\ref{eq2}) and (\ref{eq3}).}
\tablenotetext{e}{Line intensities from our observations.}
\tablenotetext{f}{Line intensities from \citet{Liu_04b}.}\\
\end{deluxetable*}

\subsection{[\ion{Fe}{3}] lines and Fe$^{++}$ abundances}

To derive the Fe$^{++}$ abundance we solved the equations of 
statistical equilibrium \citep[see, e.g.,][]{Osterbrock_06} 
for the lowest 34 levels and used the collision strengths 
of \citet{Zhang_96} and the transition probabilities of \citet{Quinet_96}. 
We calculated the Fe$^{++}$ abundance using up to 10 
[\ion{Fe}{3}] lines among the following: $\lambda$4080, $\lambda$4607, 
$\lambda$4667, $\lambda$4658, $\lambda$4701, $\lambda$4734, $\lambda$4755, 
$\lambda$4769, $\lambda$4778, $\lambda$4881, $\lambda$4986, $\lambda$5412, 
and $\lambda$5270. 
When several lines were measured, we rejected those that led to Fe$^{++}$ 
abundances much larger than the rest, since this could be due to contamination 
of these [\ion{Fe}{3}] lines with other weak lines. The final Fe$^{++}$ 
abundance is the weighted mean of all the abundances derived with the 
available lines (with weights 1/$\sigma^2$ where $\sigma$ comes from the
errors in the line intensities).  
The spectra we used from the literature have a resolution better than 2~\AA\ in
the blue range, but our observed spectra have a resolution of about 4~\AA\ and
[\ion{Fe}{3}] $\lambda4658$ is blended with \ion{O}{2} $\lambda4661$. In order
to deblend these lines we used a multiple Gaussian  profile-fitting procedure
in IC~4593, and this procedure worked properly in this object since the
Fe$^{++}$ abundance we obtain from this [\ion{Fe}{3}] line is in good agreement
with the results obtained with the other four [\ion{Fe}{3}] lines measured in
this PN.  However, the multiple Gaussian fit did not work in NGC~6210 and
NGC~2392. For NGC~6210, we corrected for the contribution of \ion{O}{2}
$\lambda4661$ using the intensity we measured for
\ion{O}{2} $\lambda\lambda$4672, 4676 lines and the expected relative 
intensities of
the \ion{O}{2} recombination lines from the same multiplet 
\citep{Peimbert_05}. We did not measure any \ion{O}{2} recombination line in
NGC~2392, but the contribution of \ion{O}{2} $\lambda4661$ to the [\ion{Fe}{3}]
$\lambda4658$ line should not be important in this object, since the Fe$^{++}$
abundance derived from [\ion{Fe}{3}] $\lambda4658$ is consistent with the
abundances obtained from  the other five [\ion{Fe}{3}] lines. In 
NGC~3587 we only measured the line [\ion{Fe}{3}] $\lambda4986$ (see 
Figure~\ref{lineas_Fe1}), which is the brightest line for the physical 
conditions of this nebula.

Three of the PNe from the literature have no identifications of 
[\ion{Fe}{3}] lines in their spectra and we used recombination lines 
measured near 4658~\AA\ (since [\ion{Fe}{3}] $\lambda4658$ should 
be the brightest line for the physical conditions of these objects) to 
derive upper limits to the Fe$^{++}$ abundance. The lines used are 
\ion{C}{4} $\lambda4659$ for IC~1747 and NGC~7026, and \ion{O}{2} 
$\lambda4661$ for IC~4406. We also calculated upper limits to the 
Fe$^{++}$ abundance for JnEr~1, where the detection of [\ion{Fe}{3}] 
lines is uncertain (see Figure~\ref{lineas_Fe2}).

Column 9 in Table~\ref{tabla_4} shows the Fe$^{++}$ abundance we obtained 
for each PN, and Column 10 shows the number of [\ion{Fe}{3}] lines that 
were used for each PN to calculate the final Fe$^{++}$ abundance. 
The error in the Fe$^{++}$ abundance derived from each line was 
calculated in the same way described above for O$^{+}$ and O$^{++}$, and
when several lines of [\ion{Fe}{3}] were measured we calculated the 
weighted mean of the Fe$^{++}$ abundances and used propagation of errors.
Two entries for NGC~6210 are given in Tables~\ref{tabla_3} and \ref{tabla_4}.
The first one shows the results obtained using our observations of this
object; the second entry shows the results we derived using the spectrum
measured by \citet{Liu_04b}. The results implied by the two spectra are very
similar. 

\subsection{Other ionization states of Fe}

As we mentioned in Section~\ref{intro}, the abundance of 
Fe$^+$ is often negligible, even for low-ionization objects.
Four of the sample PNe have measurements of [\ion{Fe}{2}] lines: 
IC~418, NGC~40, NGC~6720, and NGC~6884. Since [\ion{Fe}{2}] $\lambda$8616 
is almost insensitive to fluorescence effects \citep{Lucy_95}, we 
used this line to calculate the Fe$^+$ abundance in IC~418. The 
other three PNe had no measurements of this line and we used the 
intensity of [\ion{Fe}{2}] $\lambda7155$ and assumed the relation 
$I([\mbox{\ion{Fe}{2}]}~\lambda7155)/I([\mbox{\ion{Fe}{2}}]~\lambda8616)\sim 1$
found by \citet{Rodriguez_96} for \ion{H}{2} regions, to get an 
estimate of the Fe$^+$ abundance using the emissivities derived by 
\citet{Bautista_96}. Column 8 in Table~\ref{tabla_4} shows the 
results. All the $\mbox{Fe}^+/\mbox{H}^+$ abundances we obtained are 
lower than the values of $\mbox{Fe}^{++}/\mbox{H}^+$, with
$\mbox{Fe}^{+}/\mbox{Fe}^{++}=$ 0.21, 0.46, 0.38, and 0.18 for 
NGC~40, IC~418, NGC~6720, and NGC~6884, respectively. Although the values 
of $\mbox{Fe}^{+}$ are uncertain \citep[see ][]{Rodriguez_02}, we have 
used them to derive Fe abundances with Equation~(\ref{eq3}).
The effect of Fe$^{+}$ on the total Fe abundance is relevant only for
IC~418 and NGC~40, where Fe/H increases by 0.16 dex and 0.08 dex, respectively,
after including this ion. We expect that for 
the rest of the PNe in the sample the contribution 
of Fe$^+$ will be even less important.

Four PNe of the sample have measurements of lines from ions of higher 
ionization states than Fe$^{+3}$: 
[\ion{Fe}{5}] lines in IC~3568, NGC~6153, NGC~6210, 
NGC~6720, NGC~6826, and NGC~6884; [\ion{Fe}{6}] lines in NGC~6884; and 
[\ion{Fe}{7}] lines in IC~5217, NGC~3132, NGC~6210, and NGC~6884.
Since no atomic data are available to calculate the $\mbox{Fe}^{+4}$
abundance from optical lines, we only calculated the ionic 
abundances of $\mbox{Fe}^{+5}$ and $\mbox{Fe}^{+6}$. For
$\mbox{Fe}^{+5}$, we solved the equations of statistical equilibrium for 19
levels, using the transition probabilities and collision strengths of
\citet{Chen_99,Chen_00}. For the $\mbox{Fe}^{+6}$ calculations, we used a
model atom with 9 levels and the transition probabilities and collision 
strengths of \citet{Witthoeft_08}. The level energies in both cases
are from the compilation of \citet{Sugar_85} listed by
{\sc NIST}\footnote{http://physics.nist.gov/}, and 
the physical conditions used are the values of \T[\ion{O}{3}] and \n\ shown in 
Table~\ref{tabla_3}.

The identification of [\ion{Fe}{7}] $\lambda$6601
in IC~5217 \citep{Hyung_01a} is probably incorrect since this line 
leads to an unrealistic high abundance, 
$\mbox{Fe}^{+6}/\mbox{H}^{+}=8\times10^{-4}$, and besides, other lines 
of the same ion such as $\lambda$4990, $\lambda$5160, $\lambda$5278, and 
$\lambda$5720, which should be brighter by more than a factor of 30 
for the physical conditions of this PN, are not present in the spectra. 
In the other objects, we find
$\mbox{Fe}^{+5}/\mbox{H}^{+}= 6.0\times10^{-8}$ 
and $\mbox{Fe}^{+6}/\mbox{H}^{+} = 4.7\times10^{-8}$ for NGC~6884, 
$\mbox{Fe}^{+6}/\mbox{H}^{+} = 8.0\times10^{-8}$ for NGC~3132, and 
$\mbox{Fe}^{+6}/\mbox{H}^{+} = 1.23\times10^{-7}$ for NGC~6210. 
If we had used the emissivities of \citet{Nussbaumer_78} for 
$\mbox{Fe}^{+5}$, the derived abundances would be higher by 
factors up to 1.7. In the case of $\mbox{Fe}^{+6}$, the transition 
probabilities of \citet{Keenan_87} and the collision strengths of 
\citet{Nussbaumer_82} would lead to abundances higher by factors up 
to 1.8. 

These $\mbox{Fe}^{+5}$ and $\mbox{Fe}^{+6}$ abundances are high,
with values similar to the derived $\mbox{Fe}^{++}$ abundances, but we should
consider them with caution.
On the one hand, only one [\ion{Fe}{7}] line has been measured for these PNe, 
but we estimate that other [\ion{Fe}{7}] lines, such as $\lambda$4893,
$\lambda$5722, $\lambda$4990, or $\lambda$6089, should be brighter by factors 
between 3 and 18 for the physical conditions of these objects.
Besides, NGC~3132 and NGC~6210 can be considered low-ionization PNe, since they
have $\mbox{He}^{++}/\mbox{H}^{+}=$ 0.029 and 0.018, respectively,
and hence we do not expect that high states of 
ionization contribute much to the total abundance of Fe in these objects.
In fact, 
NGC~3132 does not have any [\ion{Fe}{5}] or [\ion{Fe}{6}] line, and 
NGC~6210 only has [\ion{Fe}{5}] $\lambda$4227. 
On the other hand, some [\ion{Fe}{6}] lines can be affected by 
fluorescence effects \citep{Chen_00}, and hence the real 
$\mbox{Fe}^{+5}$ abundance for NGC~6884 could be lower that the 
one we calculated. 

Nevertheless, we calculated what value of the total abundance of Fe 
we would get by adding all the calculated ionic abundances in these three PNe.
We used the abundances derived for Fe$^{+}$, Fe$^{++}$, Fe$^{+5}$, and
Fe$^{+6}$ and assumed that the ions we do not observe have abundances
intermediate between those derived for the observed adjacent ions. 
The results we obtain in this way are intermediate between the Fe 
abundances derived in Section~5 using Equations~(\ref{eq1}) and
(\ref{eq2})/(\ref{eq3}) except for NGC~3132, where the Fe abundance calculated 
in this way ($12+\mbox{Fe}/\mbox{H}\sim5.61$) is higher by a factor around 2.
The agreement found for the other two objects may not be significant
for the reasons mentioned above.

\subsection{Comparison with \ion{H}{2} regions}

\begin{figure}
\includegraphics[width=8.0cm,trim = 30 0 50 0,clip =yes]{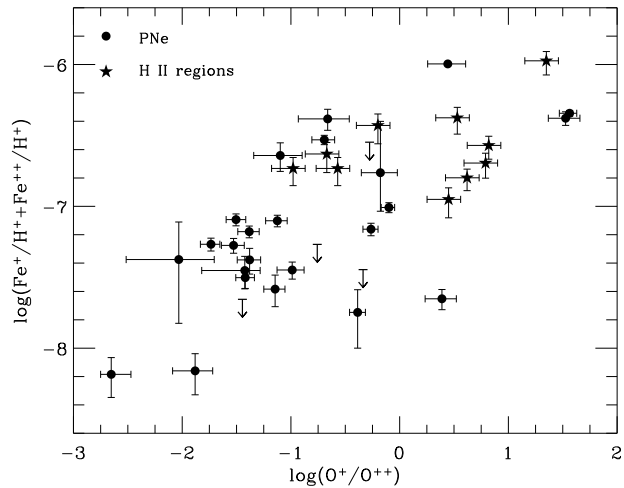}
\caption{{\small Values of $\mbox{Fe}^{+}/\mbox{H}^{+} + \mbox{Fe}^{++}/\mbox{H}^{+}$ 
as a function of the degree of ionization.}\label{Fe1}}
\end{figure}

For comparison purposes, we have selected from the literature a group of 10
Galactic \ion{H}{2} regions (see Table~\ref{tabla_5}) that have all the 
lines needed to carry out the
same analysis we have performed for the PNe. We used the values of the ionic
abundances of $\mbox{Fe}^{+}$, $\mbox{Fe}^{++}$, $\mbox{O}^{+}$, and
$\mbox{O}^{++}$ derived in the original papers since they were calculated 
using a similar procedure to the one we use for our sample
PNe. Figure~\ref{Fe1} shows the values of 
$\mbox{Fe}^{+}/\mbox{H}^{+} + \mbox{Fe}^{++}/\mbox{H}^{+}$ as a function 
of the degree of ionization (measured by the ratio
$\mbox{O}^{+}/\mbox{O}^{++}$) for all our sample objects.
The objects in Figure~\ref{Fe1}\ follow a trend of increasing 
$\mbox{Fe}^{+}/\mbox{H}^{+} + \mbox{Fe}^{++}/\mbox{H}^{+}$ for
decreasing degree of ionization. This trend reflects that for lower 
ionization objects Fe$^{++}$ becomes more important in the total Fe 
abundance.

\begin{deluxetable}{lccc}
\tabletypesize{\small}
\tablecaption{Total abundances for the sample of \ion{H}{2} regions: \{X\} = 12 + $\log$ (X/H) \label{tabla_5}}
\tablewidth{0pt}
\tablehead{
\colhead{Object} & \colhead{\{Fe\}\tablenotemark{a}} &
\colhead{\{Fe\}\tablenotemark{b}} & \colhead{Ref.}
}
\startdata
\objectname{M8}  & 5.69$^{+0.12}_{-0.23}$& 5.62$^{+0.11}_{-0.21}$ & 1\\
\objectname{M16} & 5.16$^{+0.12}_{-0.23}$& 5.20$^{+0.11}_{-0.19}$ & 2\\	
\objectname{M17} & 5.82$^{+0.12}_{-0.23}$& 5.62$^{+0.15}_{-0.32}$ & 1\\
\objectname{M20} & 5.31$^{+0.12}_{-0.24}$& 5.31$^{+0.12}_{-0.21}$ & 2\\
\objectname{M42} & 6.02$^{+0.12}_{-0.23}$& 5.78$^{+0.15}_{-0.33}$ & 3\\
\objectname{M43} & 6.01$^{+0.13}_{-0.24}$& 6.03$^{+0.12}_{-0.23}$ & 4\\
\objectname{NGC 3576} & 5.92$^{+0.12}_{-0.22}$& 5.91$^{+0.14}_{-0.30}$ & 5\\
\objectname{NGC 3603} & 6.14$^{+0.13}_{-0.24}$& 5.74$^{+0.16}_{-0.36}$ & 2\\
\objectname{NGC 7635} & 5.40$^{+0.12}_{-0.23}$& 5.43$^{+0.11}_{-0.21}$ & 4\\
\objectname{S311}    &  5.17$^{+0.12}_{-0.23}$& 5.05$^{+0.12}_{-0.21}$ & 6\\
\enddata
\tablenotetext{a}{Derived using the ICF of equation~(\ref{eq1}).}
\tablenotetext{b}{Derived using the ICF of equations~(\ref{eq2}) and (\ref{eq3}).} 
\tablerefs{(1) \citet{GarciaRojas_07}, (2) \citet{GarciaRojas_06}, (3) \citet{Esteban_04}, 
 (4) \citet{Rodriguez_02}, (5) \citet{GarciaRojas_04}, (6) \citet{GarciaRojas_05}.}\\
\end{deluxetable}

\section{Total abundances \label{total}}

We calculate the total oxygen and iron abundances as 
$\mbox{O/H}=
(\mbox{O}^{+}/\mbox{H}^{+}+\mbox{O}^{++}/\mbox{H}^{+})\times\mbox{ICF(O)}$ and
$\mbox{Fe/H}=(\mbox{O/H})\times(\mbox{Fe}^{++}/\mbox{O}^{+})\times\mbox{ICF(Fe)}$. 
The ICFs used for Fe are those from Equations~(\ref{eq1}) and
(\ref{eq2})/(\ref{eq3}); the ICF scheme from 
\citet{Kingsburgh_94} has been adopted for O:
\begin{equation}\label{eq4}
\mbox{ICF(O)}=
\left(\frac{\mbox{He}^{+} + \mbox{He}^{++}}{\mbox{He}^{+}}\right)^{2/3}.
\end{equation}

Column 6 in Table~\ref{tabla_4} shows the values of ICF(O) for 
each PN of the sample. This ICF accounts for the contribution of ions of 
higher degree of ionization than O$^{++}$, and its value is $\sim1.0$ for
most of our low-ionization PNe and for the \ion{H}{2} regions. 
The largest values are 1.18 and 1.23 for NGC~3242 and NGC~2392, 
respectively, the sample PNe with the highest values of
$I(\mbox{\ion{He}{2}}~\lambda4686)/I(\mbox{H}\beta)$, 0.26 and 0.33,
respectively. The low abundance of O$^{+3}$ is confirmed for some of
the PNe (IC~3568, NGC~40, NGC~6720, and NGC~6884), where the O$^{+3}$
abundances calculated by \citet{Liu_04a} using the [\ion{O}{4}] line at
$25.9~\mu$m, contribute less than 10\% to the total
abundances (in fact, these PNe were among the ones we used to define the
criteria for the sample selection -- see Section~\ref{sample}).
Column 7 in Table~\ref{tabla_4} shows 
the final values of the O abundances for the sample PNe.
Our values of the total O abundances are in reasonable agreement 
with previous values in the literature. The highest discrepancies are 
found for Cn~3-1 and NGC~2392. 
The value of the O abundance in NGC~2392 ranges from
$12+\log(\mbox{O}/\mbox{H})=8.35$ to 8.61 according to different authors
\citep{Barker_91,Henry_00,Pottasch_08}, and we find
$12+\log(\mbox{O}/\mbox{H})=8.25$.
For Cn~3-1, the difference of $\sim$0.2 dex with the results of 
\citet{Wesson_05} is due to the higher electron density we used.

The Fe abundances obtained from two of the ICFs presented in
Section~\ref{intro} are shown in Tables~\ref{tabla_4} and \ref{tabla_5} 
for the sample PNe and the group of \ion{H}{2} regions. The first 
of the listed values is based on the ICF implied by photoionization 
models; the second is based on the ICF derived using those objects with
measurements  of both [\ion{Fe}{3}] and [\ion{Fe}{4}] lines.
Lowering the first values by $\sim 0.3$~dex we get a third value for Fe/H
and, as discussed in Section~\ref{intro}, these three values of
Fe/H can be used to constrain the real value of the gaseous Fe abundance.

The values of Fe$^{++}$/Fe are very low for the high excitation 
objects, with the lowest value found for IC~3568, the PNe with 
the highest degree of ionization in the sample, where Fe$^{++}$ 
contributes less than 10\% to the derived Fe abundance. The fact 
that both Fe$^{++}$ and O$^+$ have very low abundances in the nebulae 
with high degrees of ionization suggest that for those objects the
real ICFs can show a high dispersion at a given value of
$\mbox{O}^{+}/\mbox{O}^{++}$ \citep[see Figure 2 in][]{Rodriguez_05}.
This dispersion will translate into errors in the Fe abundances derived 
from Equations~(\ref{eq1}) and (\ref{eq2})/(\ref{eq3}).
Since these are also the objects where Equations~(\ref{eq1})
and (\ref{eq2}) lead to the most discrepant results, their
Fe abundances are the less well constrained.

For 11 PNe of the 28 in our sample, there are previous calculations of the Fe
abundance, and the differences between our values and the values obtained by
other authors go up to 1.7 dex. These differences are due to the use of
different atomic data and ICFs. We are using what we think are the best values
for all the atomic data \citep[see][]{Rodriguez_02, Rodriguez_05} and our
detailed analysis and the fact that we are using the same procedure for a
relatively large sample of objects allow us to compare the results for
different objects and to draw some inferences.

\begin{figure}
\subfigure{
\includegraphics[width=9cm,trim = 40 40 0 0,clip =yes]{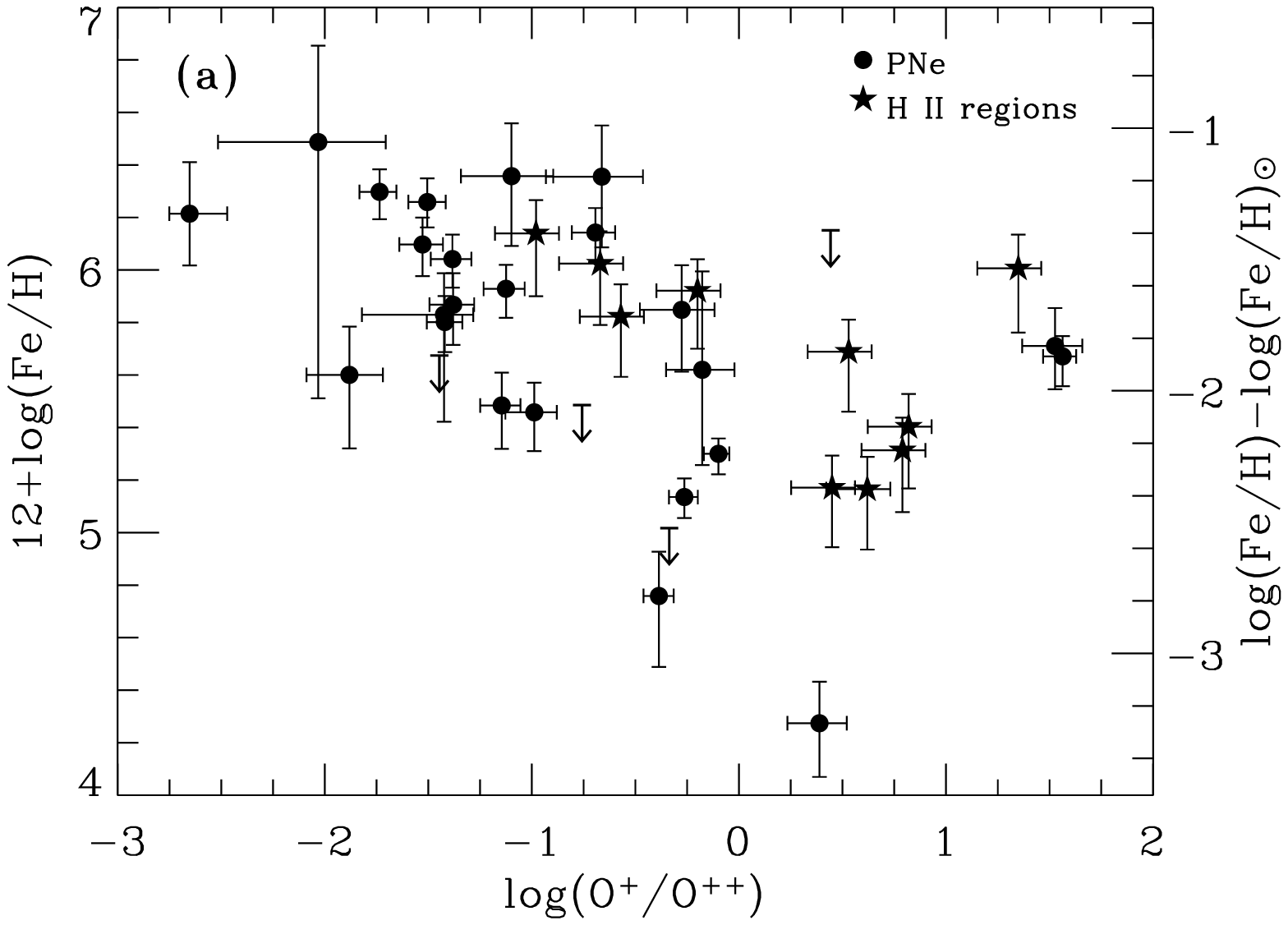}}
\subfigure{
\includegraphics[width=9cm,trim = 40 20 0 0,clip =yes]{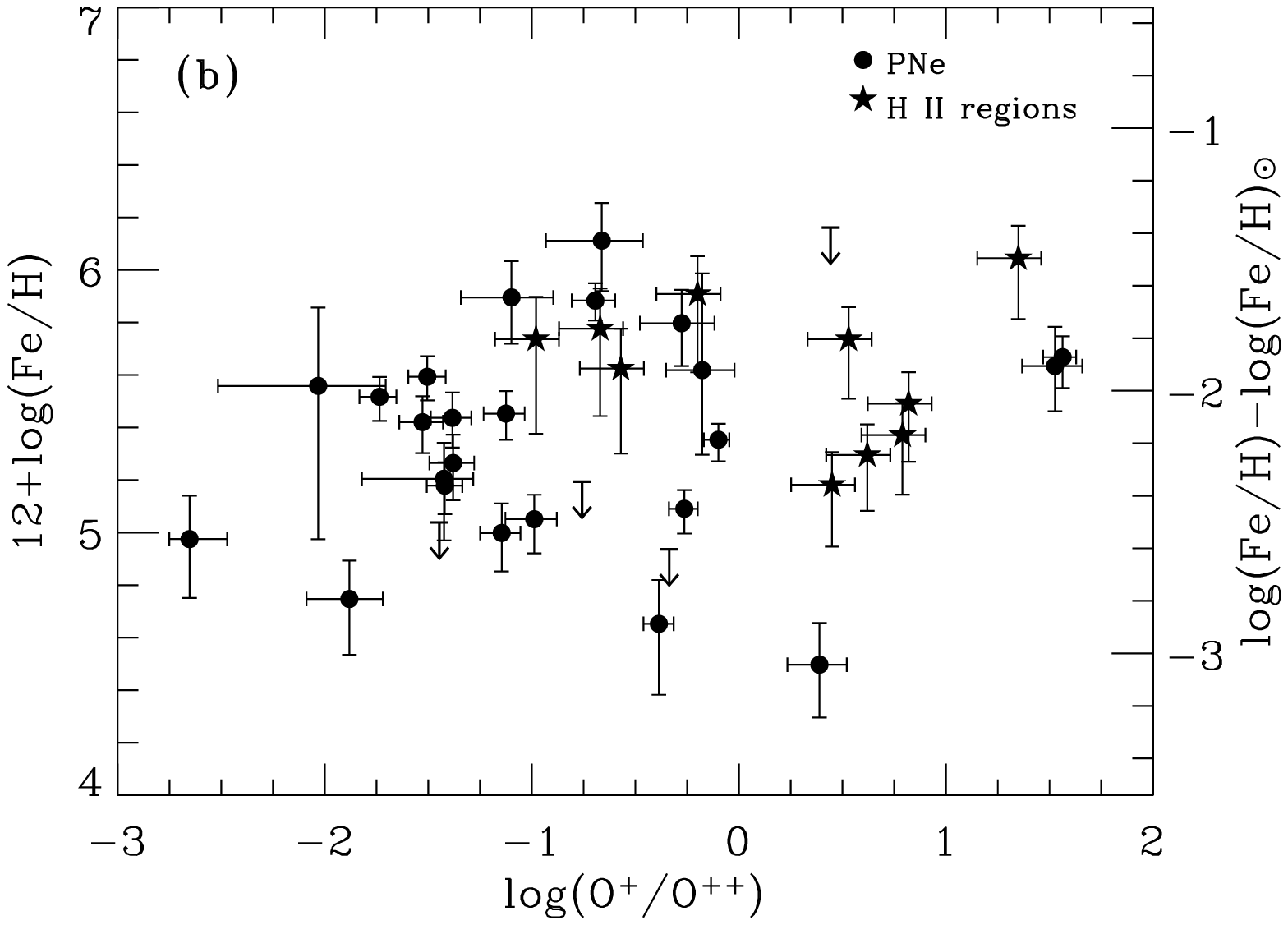}}\\
\caption{{\small Values of Fe/H (left axis) and the depletion factors 
for Fe/H ($[\mbox{Fe/H}]= \log(\mbox{Fe/H}) - \log(\mbox{Fe/H})_{\odot}$, right 
axis) as a function of the degree of ionization.
Panel (a) shows the values obtained  from equation~(\ref{eq1}), and panel (b)
is from equations~(\ref{eq2})/(\ref{eq3}).}\label{FeH1}}
\end{figure}

Figure~\ref{FeH1} shows the values of the Fe abundance derived with the 
two ICFs from Equations~(\ref{eq1}) and (\ref{eq2})/(\ref{eq3}) as a function
of the degree of ionization for all our objects.
The axes at the right show the depletion factor,
$[\mbox{Fe}/\mbox{H}]=\log(\mbox{Fe/H})-\log(\mbox{Fe/H})_{\odot}$, with
$12+\log(\mbox{Fe/H})_{\odot}=7.54\pm0.03$ \citep{Lodders_03}.
It can be seen that all the objects in our sample (both PNe and \ion{H}{2}
regions) have depletion factors below $-1.1$ dex. Hence, less than 10\% 
of the solar Fe abundance is present in the gas.
If the solar abundance can be considered a good reference for the
total Fe abundance (in gas and dust) of these objects (see
Section~7), this implies that more than 90\% of the Fe 
atoms are  deposited onto dust grains. 

The Fe abundances in Figure~\ref{FeH1}(a) show a trend that suggests that objects
with higher degree of ionization have somewhat lower depletion factors.
This trend could be related to dust destruction in PNe with harsh
radiation fields.
However, the trend is not present in Figure~\ref{FeH1}(b), and if the discrepancy
in the Fe abundances implied by Equations~(\ref{eq1}) and (\ref{eq2}) is due to
a combination of errors in different atomic data, the depletion 
factors will be intermediate between the extreme values we are considering, and
the trend is likely to disappear.

\section{PNe with high density and age dependence \label{highdensity}}

We have analyzed five additional high-density PNe (see Table~\ref{tabla_6})
 because they are likely to be younger than the PNe in our sample,
since they have high $\mbox{H}\beta$ surface brightness
($S(\mbox{H}\beta)>5\times10^{-13}$ erg s$^{-1}$ cm$^{-2}$ arcsec $^{-2}$)
and high electron densities (\n\ $>$ 25,000 \cm). 
To calculate $S(\mbox{H}\beta)$ we used the total 
$\mbox{H}\beta$ fluxes and the visual extinction coefficient 
from \citet{Cahn_92}, and the angular sizes from 
\citet{Acker_92} and \citet{Tylenda_03}.
We did not include these objects in the original 
sample since they show important differences in the 
values obtained with different diagnostic line ratios 
(see Table~\ref{tabla_6}), which complicate the calculation of 
physical conditions and abundances.
We decided to constrain the values of the abundances in these
PNe by analyzing them in three different ways.
The first two rows for the results of each PN in Table~\ref{tabla_6}
show the \T's and ionic abundances derived separately from the two
values of \n\ we had for each PN, whereas the third row shows
the \T's and ionic abundances obtained by assuming that there is a
density gradient in each object and considering different values of \n\
for each ion according to their ionization potential. 
The differences in the Fe abundances obtained with these three procedures 
go up to 0.9 dex, and this gives us an idea of the uncertainties involved.

\begin{deluxetable*}{ccccccccccccc}
\tabletypesize{\small}
\tablecaption{Physical Conditions, Ionic and Total Abundances of high density PNe. 
\{X$^{+i}$\} = 12 + $\log$ (X$^{+i}$/H$^{+}$), \{X\} = 12 + $\log$ (X/H) \label{tabla_6}}
\tablewidth{0pt}
\tablehead{
\colhead{\n} & \colhead{\TN} & \colhead{\TO} & \colhead{\{He$^{+}$\}} & \colhead{\{He$^{++}$\}} & \colhead{\{O$^{+}$\}} 
& \colhead{\{O$^{++}$\}} & \colhead{ICF} & \colhead{\{O\}} & \colhead{\{Fe$^{++}$\}} & \colhead{N\tablenotemark{a}} 
& \colhead{\{Fe\}\tablenotemark{b}}  & \colhead{\{Fe\}\tablenotemark{c}}\\
(\cm) & (K) & (K) & & & & & & & & & & }
\startdata
\multicolumn{13}{c}{\normalsize {\objectname{M 1-74} {\small(052.2-04.0\tablenotemark{d})}}} \\
\n[\ion{S}{2}]\,=\,22300 & 12100  & 9800 & 11.00 & 8.07 & 7.23 & 8.56 & 1.00 & 8.58 & 5.27 & 3 & 6.47 & 5.89\\
\n[\ion{Ar}{4}]\,=\,78500 & 8100     & 9200 & 11.03 & 9.00 & 8.48 & 8.70 & 1.01 & 8.90 & 5.82 & 3 & 6.18 & 6.16\\
\n[\ion{S}{2}]\,and\,\n[\ion{Ar}{4}] & 12100  & 9200 & 11.03 & 9.00 & 7.23 & 8.70 & 1.01 & 8.71 & 5.27 & 3 & 6.59 & 5.95\\
\multicolumn{13}{c}{\normalsize {\objectname{Me 2-2} {\small(100.0-08.7\tablenotemark{d})}}} \\
\n[\ion{S}{2}]\,=\,1000 & 15000  & 10900 & 11.18 & 8.09 & 6.33 & 8.23 & 1.00 & 8.23 & 4.59 & 3 & 6.29 & 5.43\\
\n[\ion{Ar}{4}]\,=\,34500 & 9500     & 10400 & 11.17 & 8.09 & 7.79 & 8.31 & 1.00 & 8.43 & 4.95 & 3 & 5.51 & 5.33\\
\n[\ion{S}{2}]\,and\,\n[\ion{Ar}{4}] & 15000  & 10400 & 11.17 & 8.09 & 6.33 & 8.31 & 1.00 & 8.32 & 4.59 & 3 & 6.37 & 5.46\\
\multicolumn{13}{c}{\normalsize {\objectname{NGC 5315} {\small(309.1-04.3\tablenotemark{d})}}}\\
\n[\ion{S}{2}]\,=\,8400  & 10800  & 9200 & 11.09 & 7.62 & 7.57 & 8.57 & 1.00 & 8.61 & 3.95 & 3 & 4.86 & 4.45\\
\n[\ion{Cl}{3}]\,=\,28600  & 8600     & 8900 & 11.09 & 7.62 & 8.38 & 8.63 & 1.00 & 8.83 & 4.58 & 3 & 4.96 & 4.92\\
\n[\ion{S}{2}]\,and\,\n[\ion{Cl}{3}] & 10800  & 8900 & 11.09 & 7.62 & 7.57 & 8.63 & 1.00& 8.67 & 3.95 & 3 & 4.91 & 4.47\\
\multicolumn{13}{c}{\normalsize {\objectname{NGC 6790} {\small(037.8-06.3\tablenotemark{d})}}}\\
\n[\ion{Cl}{3}]\,=\,25700  & 14800  & 13000 & 11.00 & 9.52 & 7.64 & 8.43 & 1.02 & 8.50 & 4.54 & 6 & 5.29 & 4.98\\
\n[\ion{Ar}{4}]\,=\,139500  & 7800     & 11300 & 10.99 & 9.52 & 8.68 & 8.60 & 1.02 & 8.95 & 5.41 & 6 & 5.64 & 5.68\\
\n[\ion{Cl}{3}]\,and\,\n[\ion{Ar}{4}] & 14800  & 11300 & 10.99 & 9.52 & 7.64 & 8.60 & 1.02 & 8.65 & 4.54 & 6 & 5.43 & 5.03\\
\multicolumn{13}{c}{\normalsize {\objectname{NGC 6807} {\small(042.9-06.9\tablenotemark{d})}}}\\
\n[\ion{S}{2}]\,=\,15800 & 14600  & 10600 & 10.35 & 8.45 & 6.52 & 8.55 & 1.01 & 8.56 & 4.74 & 3 & 6.56 & 5.64\\
\n[\ion{Ar}{4}]\,=\,54600 & 9800  & 10100 & 10.35 & 8.45 & 7.60 & 8.65 & 1.01 & 8.69 & 5.17 & 3 & 6.13 & 5.69\\
\n[\ion{S}{2}]\,and\,\n[\ion{Ar}{4}] & 14600  & 10100 & 10.36 & 8.45 & 6.52 & 8.65 & 1.01 & 8.65 & 4.74 & 3 & 6.65 & 5.68
\enddata
\tablerefs{{\it Line intensities from:} \citet{Wesson_05} for M~1-74, Me~2-2, and NGC~6807; \citet{Tsamis_03} for 
NGC~6790; and \citet{Liu_04b} for NGC~5315.}
\tablenotetext{a}{Number of [\ion{Fe}{3}] lines used in the Fe$^{++}$ abundance
determination.}
\tablenotetext{b}{Derived using the ICF of equation~(\ref{eq1}).}
\tablenotetext{c}{Derived using the ICF of equation~(\ref{eq2}) and (\ref{eq3}).}
\tablenotetext{d}{PNG identifications from \citet{Acker_92}.}
\end{deluxetable*}

The depletion factors of these five PNe are similar to the 
values found for the original sample of 28 PNe, with 
more than 87\% of their Fe atoms deposited onto dust grains.
We do not find any significant 
correlation between our derived Fe abundances and $S(\mbox{H}\beta)$ 
(or \n) in the whole group of 33 PNe, which includes objects that are 
likely to be old such as JnEr~1 or NGC~3587, with 
$S(\mbox{H}\beta) = 8.3\times10^{-17}$ and 
$1.9\times10^{-15}$ erg s$^{-1}$ cm$^{-2}$ arcsec$^{-2}$, respectively, or
objects that are likely to be young, like MyCn~18 with
$S(\mbox{H}\beta) = 4.4\times10^{-11}$ erg s$^{-1}$ cm$^{-2}$ arcsec$^{-2}$ and
the high-density objects. 
The low Fe abundances of our sample PNe and the lack of correlation
of the Fe abundances with parameters that can be related to
the ages of the objects suggest that no significant destruction of
refractory dust grains has taken place in these objects, in agreement
with the results found by \citet{Stasinska_99}.
However, our sample objects are relatively bright PNe, where weak 
lines have been measured. Hence, we must be missing in our sample
the oldest PNe, characterized by low densities, low surface brightness,
and large nebular radii. In principle, these old objects could
show some evidence for dust destruction.

\section{Discussion \label{discussion}}

The O abundances for the original sample of 28 PNe and for the 
whole sample of 33 PNe are in the ranges 8.25--8.81 and 8.23--8.95 
respectively, whereas for the \ion{H}{2} regions the range is 
8.39--8.56.
The dispersion is much higher for the sample PNe, and it can be due
to (1) the production or destruction of oxygen in the progenitors of the
PNe, or (2) different chemical compositions of the clouds where the
central stars formed. In the latter case, different values of Fe/H
should be used as the reference abundance to determine depletions.

Both models and observations indicate that oxygen can be produced or
destroyed in low-metallicity PNe 
\citep[i.e., for Magellanic Clouds metallicities, see, e.g.,][]
{Karakas_03T, Marigo_03, Leisy_06, Karakas_08}, but the standard 
theoretical results do not predict any significant change in the 
oxygen abundance at higher metallicities 
\citep{Marigo_03, Karakas_03, Karakas_08}. However, several authors 
do not consider this a settled issue 
\citep[see][]{Perinotto_98, Pottasch_06, Karakas_08} and in principle,
the O abundances of our sample PNe could still be affected by 
these effects.

On the other hand, if the dispersion in the O abundances (or part of it) were 
due to the different initial compositions of the central stars,
the Fe/O ratio, that changes slowly with metallicity, would be better
suited than Fe/H to derive depletion factors.
However, the Fe/O ratio would need a correction for the depletion
of oxygen in dust grains.
A dust-phase oxygen abundance of 180 parts per million \citep{Cardelli_96}
represents 30\% of the solar abundance ($12+\log(\mbox{O/H})_{\odot}=8.76$
\citet{Lodders_03}). Assuming that this 30\% also applies to our sample PNe,
and using the solar value of Fe/O, $\log(\mbox{Fe/O})_{\odot} = -1.22$
\citep{Lodders_03}, the depletion factors implied by Fe/O and Fe/H are
similar: the ranges that take into account all the possible values
go from $-3.6$ to $-0.8$ for $\log(\mbox{Fe/O})$,
and from $-3.6$ to $-1.1$ for $\log(\mbox{Fe/H})$. 

Some H-poor central stars of PNe show an iron deficiency 
of up to $\sim$1--2 dex, which has been associated with 
\emph{s}-process nucleosynthesis occurring in the intershell during 
the AGB or post-AGB phase 
\citep[][and references therein]{Miksa_02,Werner_03,Werner_06, Karakas_08}.
In these PNe with H-poor central stars some Fe could be depleted 
because of this process, and 13 PNe of our sample have this type of
central star: Cn~3-1, IC~1747, IC~5217, JnEr~1, M~1-73, 
MyCn~18, NGC~40, NGC~5315, 
NGC~6153, NGC~6543, NGC~6572, NGC~6803, and NGC~7026 
\citep{Hucht_81, Liebert_88, Mendez_91, Tylenda_93, Crowther_98, 
Parthasarathy_98, Liu_00, Bohigas_01, Lee_07, Sterling_08}.
However, \citet{Sterling_08} concluded recently that the compositions
of PNe with H-poor central stars are not different from those PNe
that have H-rich central stars. 
Furthermore, we do not see any systematic difference in the Fe 
abundances derived for PNe with H-rich and H-poor central stars. 

The error bars in Figure~\ref{FeH1} do not take into account 
those uncertainties arising from the ICF, and these errors could 
explain most of the dispersion in Fe/H shown by the high 
excitation objects. 
However, some of the dispersion shown by all the PNe is likely
to be real and due to different depletion factors,
since it is quite high in some cases, with differences
in the Fe abundances up to $\sim1.5$ dex for those objects with
$\log(\mbox{O}^{+}/\mbox{O}^{++})\sim-0.5$. 
We did not find any obvious difference in the morphology, the 
type of the central star, or the dust chemistry of
the PNe with the highest and lowest depletion factors. 
Using the $\mbox{C/O}$ values and the infrared dust features 
identified in the literature, we infer that 13 PNe of the sample 
are O-rich ($\mbox{C/O} < 1$), 8 are C-rich ($\mbox{C/O} > 1$ and/or
they show PAHs or SiC in their spectra), and 7 are uncertain since
$\mbox{C/O} \sim 1$ or they do not have clear dust features 
\citep{Cohen_86, Rola_94, Roche_96, Kholtygin_98, Kwitter_98, 
Pottasch_99, Henry_00, Liu_00, Casassus_01, Tsamis_03, 
Liu_04a, Pottasch_04, Tsamis_04, Cohen_05, Wesson_05, Smith_08}.
It is not clear yet which are the main iron compounds condensing
onto dust grains, but O-rich environments are expected to have metallic
iron grains, silicates, and oxides, whereas C-rich environments can have
their iron in the form of metallic grains, Fe$_3$C, FeSi,
FeS, and FeS$_2$ \citep{Whittet_03, Ferrarotti_06}.
The fact that we do not see any 
systematic difference between the Fe abundances of C-rich and 
O-rich PNe suggests that the iron depletion efficiencies in C-rich and 
O-rich environments are similar. This is supported by
the lack of systematic differences between the iron 
abundances of PNe with H-rich and H-poor central stars mentioned above, 
since H-poor central stars are usually C-rich \citep{deMarco_01, Pena_03}.

\subsection{Dust-to-gas ratios \label{ratio}}

If we assume that gas and dust evolve together, we can derive a lower
limit to the dust-to-gas mass ratio taking into account that at least
90\% of the solar Fe abundance is deposited onto dust grains in the
objects of our sample. We find that
$M_{\rm dust}/M_{\rm gas}\geq1.3\times10^{-3}$, but elements like
Si and Mg can have similar contributions to the mass of dust grains
and the contributions of C and O can be even higher 
\citep[see e.g.][]{Sofia_94}.
The dust-to-gas mass ratios derived by \citet{Stasinska_99} are 
lower than this lower limit by factors up to 17 for 10 PNe of our 
sample. This illustrates how difficult it is to find a reliable 
value for the dust-to-gas ratio using infrared dust emission 
\citep[see also][]{Stasinska_99, Gorny_01, Phillips_07}.

\section{Summary and Conclusions}

We have constrained the iron abundances in a sample of 33 
low-ionization PNe using the ICFs scheme developed by \citet{Rodriguez_05}. 
This is the largest sample of PNe where the iron abundance has been 
calculated, including 18 nebulae with first determinations.
The fact that we considered quite drastic changes in 
the atomic data involved, and the fact that we analyzed the whole 
sample with the same procedure, allow us to constrain the real value of 
the Fe abundances and compare the results between objects. The depletion 
factors ($[\mbox{Fe/H}] = \log(\mbox{Fe/H}) - \log(\mbox{Fe/H})_{\odot}$) 
of the PNe are below $-0.9$ dex, which implies that more than $\sim$90\% 
of the total Fe abundance is condensed onto dust grains. This result 
suggests that the efficiency with which Fe atoms attach to dust grains 
is higher that the one predicted by the models of AGB dust production 
of \citet{Ferrarotti_06}. We derived a lower limit to the dust-to-gas mass 
ratio for the sample PNe just by considering that 90\% of their iron 
atoms are condensed onto dust grains,
$M_{\rm dust}/M_{\rm gas}\geq1.3\times10^{-3}$. Other elements,
such as Si, Mg, C, and O, will have similar or higher contributions to 
the mass of dust grains.

The derived depletion factors span about two orders of magnitude, but
we have not found any significant correlation between the derived Fe 
abundances and the evolutionary parameters of our sample PNe (surface 
brightness or electron density), in agreement with the results 
obtained by \citet{Stasinska_99}. This result suggests that no
significant destruction of dust grains is taking place in these objects.
The different depletion factors shown by the PNe could be due either
to different dust condensation efficiencies or to the destruction of a 
minor dust component in some objects. 

We do not find any systematic difference in the Fe abundances
that can be related to the morphology, the type of the central star 
(i.e., H-rich or H-poor central star), or the dust chemistry. 
This result suggests that C-rich and O-rich progenitor stars
have similar iron depletion efficiencies.
However, the lack of identifications of dust features in most of 
the sample PNe, and the uncertainties associated with the value of 
the $\mbox{C/O}$ ratio, imply that this issue cannot be considered 
settled.

We have compared our results with the values derived for a group of 
10 \ion{H}{2} regions using the same procedure. The high depletion factors
found for both kinds of objects imply that the atmospheres of AGB stars 
deplete refractory elements onto dust grains as efficiently as
molecular clouds, whereas dust destruction processes are not very
efficient in either \ion{H}{2} regions or PNe.

\begin{acknowledgements}
The authors thank Jorge Garc\'ia-Rojas for helpful comments, 
Roger Wesson for sending some extra measurements of line 
intensities, and Michael Witthoeft for providing the atomic data 
for [\ion{Fe}{7}].
We also thank an anonymous referee for helpful comments that improved
the paper.
This work has made use of NASA's Astrophysics Data 
System, and the SIMBAD database operated at CDS, Strasbourg, France.
GD-I and MR acknowledge support from Mexican CONACYT project 50359-F.
\end{acknowledgements}

\bibliographystyle{apj}

\end{document}